%% file: paper.tex
\documentclass[amsthm,10pt]{elsart}
\usepackage{amsmath,amsthm,amsfonts,amssymb}
\usepackage[dvips]{graphicx}

\newcommand{\bra}{\bigl\langle}
\newcommand{\ket}{\bigr\rangle}
\newcommand{\bbra}{\langle\!\langle}
\newcommand{\kket}{\rangle\!\rangle}

\newcommand{\psibar}{\bar{\psi}}
\newcommand{\bQ}{\overline{Q}}
\newcommand{\Op}{{\mathcal{O}}}
\renewcommand{\P}{{\mathcal{P}}}

\newcommand{\Nev}{N_{\text{ev}}}

\newcommand{\xv}{\vec{x}}
\newcommand{\Tr}{\operatorname{Tr}}

\newlength{\colw}
\journal{Computer Physics Communications}

\begin{document}
\begin{frontmatter}

\title{Practical all-to-all propagators for lattice QCD}

\author{Justin Foley},
\author{K. Jimmy Juge},
\ead{juge@maths.tcd.ie}
\author{Alan \'O Cais},
\ead{alanc@maths.tcd.ie}
\author{Mike Peardon},
\author{Sin\'ead M. Ryan},
\author{Jon-Ivar Skullerud\corauthref{cor}}
\address{}\vspace{-8mm}
\author{(TrinLat Collaboration)}
\address{School of Mathematics, Trinity College, Dublin 2, Ireland.}
\corauth[cor]{Corresponding author. Email: jonivar@maths.tcd.ie}


\begin{abstract}
A new method for computing all elements of the lattice quark propagator is
proposed. The method combines the spectral decomposition of the
propagator, computing the lowest eigenmodes exactly, with noisy
estimators which are `diluted', i.e. taken to have support only on a
subset of time, space, spin or colour.  We find that the errors are
dramatically reduced compared to traditional noisy estimator techniques.
\end{abstract}

\begin{keyword}
Lattice QCD, Spectral decomposition, Stochastic estimators, Variance
reduction
\PACS 12.38 Gc
\end{keyword}

\end{frontmatter}

\section{Introduction}
\label{sec:intro}
\input{intro.tex}
\section{Theory and notation}
\label{sec:theory}
\input{dilute.tex}

\section{Implementation for QCD correlators}
\label{sec:qcd_impl}
\input{implement.tex}

\section{Results}
\label{sec:results}
\input{results.tex}
\section{Conclusions}
\label{sec:concl}
\input{conclude.tex}

\section*{Acknowledgments}
This work was funded by an IRCSET award SC/03/393Y and the IITAC PRTLI
initiative.


\input{paper.bbl}
\end{document}

%% file: intro.tex
Hadron spectroscopy has traditionally been performed in lattice QCD by
computing quark propagators from one or a few points on the lattice
(usually the origin) to all other points --- so-called point or
point-to-all propagators --- by inverting the fermion matrix with a
point source, and combining the resulting propagators with appropriate
operators to produce the desired hadronic correlators.  This method
does not require massive computing power, but restricts
the accessible physics to primarily the flavour non-singlet spectrum.
Flavour singlet mesons, as well as condensates and other quantities
containing quark loops, require propagators with sources everywhere in
space, and other methods, notably noisy sources, have been used for
this purpose.

Furthermore, point propagators restrict the interpolating operator
basis used, for example, to produce early plateaux in effective
masses, since a new inversion must be performed for every operator
that is not restricted to a single lattice point.

Point propagators also throw away a large portion of the
information contained in the gauge configurations.  It is likely that
a limited number of expensive configurations with light dynamical
fermions will be available in the near future, and it
will be highly desirable to extract as much information as possible from
these lattices.

All-to-all propagators \cite{Bitar:1988bb,Kuramashi:1993ka,Dong:1993pk,deDivitiis:1996qx,Eicker:1996gk,Michael:1998sg,McNeile:2000xx,Wilcox:1999ab,Neff:2001zr,Duncan:2001ta,DeGrand:2002gm,Bali:2005fu,Peardon:2002ye} provide a solution to these problems, but are
usually too expensive to compute exactly as this requires an unrealistic
number of quark inversions. Stochastic estimates tend to be very noisy
and variance reduction techniques are crucial in order to separate the
signal from the noise.  In this paper we propose an exact algorithm to
compute the all-to-all propagator utilising the idea of low-mode
dominance corrected by a stochastic estimator which yields the exact
all-to-all propagator in a finite number of quark inversions.
Some preliminary results were presented in Ref.~\cite{O'Cais:2004ww}.

The structure of the paper is as follows:  In Sec.~\ref{sec:theory} we
describe the principles of the method, and the notation we will use.
Sec.~\ref{sec:qcd_impl} describes how QCD correlators, and meson
two-point functions in particular, are computed in this framework.  In
Sec.~\ref{sec:results} we present results for the meson spectrum,
including P-waves and static--light mesons, using this
method, and compare these with results using traditional methods.
Finally, in Sec.~\ref{sec:concl} we present our conclusions.

%% file: dilute.tex


In the following, latin indices $i,j,\ldots$ denote colour; greek
indices $\alpha,\beta,\ldots$ denote spin. Indices in brackets denote
eigenvectors and dilution indices, while indices in square brackets
denote independent noise vectors; all of these will be introduced in
the following.  We use $\bra u,v\ket$ to denote the inner product (dot
product) of two fermion vectors $u$ and $v$, 
\begin{equation}
\bra u(t),v(t)\ket
 \equiv \sum_{\xv,\alpha,i} u^{i\alpha}(\xv,t)^* v^{i\alpha}(\xv,t)\,.
\end{equation}
If the $t$-argument is absent from either of the vectors, the
product is taken to be a global dot product, i.e.\ summed over all
$t$.

\subsection{Noisy Estimators and Dilution}
The standard method of estimating the all-to-all quark propagator is
by sampling the vector space stochastically. One generates an ensemble
of random, independent noise vectors,
$\{\eta_{[1]},\cdots,\eta_{[N_r]}\}$, with the property
\begin{equation}\label{eq:mixing}
\bbra\eta_{[r]}(x)\!\otimes\!\eta_{[r]}(y)^\dag\kket=\delta_{x,y}\,,
\end{equation}
where $\bbra\cdots\kket$ is the expectation value over the
distribution of noise vectors.  Each component of the noise vectors
has modulus 1,
\begin{equation}
\eta^{i\alpha}(x)^*\eta^{i\alpha}(x) = 1 \quad\text{(no summation)}\,.
\label{eq:mod1}
\end{equation}
The solution vectors $\psi_{[r]}$ are obtained in the usual way,
\begin{equation}
\psi_{[r]}(x)=M^{-1}\eta_{[r]}(y)\,.
\end{equation}
The quark propagator from any point $x$ to any other point $y$ is given by
\begin{equation}
M^{-1}(y,x)^{ij}_{\alpha\beta}
 =\bbra\psi_{[r]}\!\otimes\!\eta^\dagger_{[r]}\kket^{ij}_{\alpha\beta} 
 =\lim_{N_r\rightarrow\infty}\frac{1}{N_r}\sum_r^{N_r}\psi^{i\alpha}_{[r]}(y)\eta^{j\beta}_{[r]}(x)^\dag\,.
\label{eq:noisy}
\end{equation}
This method is noisy because it relies on delicate cancellations in
the $\Op(1)$ noise over many
samples to find the signal, which falls off exponentially with the
separation. We propose to remove the $\Op(1)$ random noise by
``diluting'' the noise vector in some set of variables $(j)$ such that
$\eta=\sum_j\eta^{(j)}$,
resulting in an substantial reduction in the variance. A particularly
important example of dilution for measuring temporal correlations in
hadronic quantities is ``time dilution" where the noise vector is
broken up into pieces which only have support on a single timeslice
each,
\begin{equation}
\eta(\xv,t)=\sum_{j=0}^{Nt-1}\eta^{(j)}(\xv,t) \, ,
\label{eq:timedilute}
\end{equation}
where $\eta^{(j)}(\xv,t)=0$ unless $t=j$.

\begin{figure}
\includegraphics*[width=\colw]{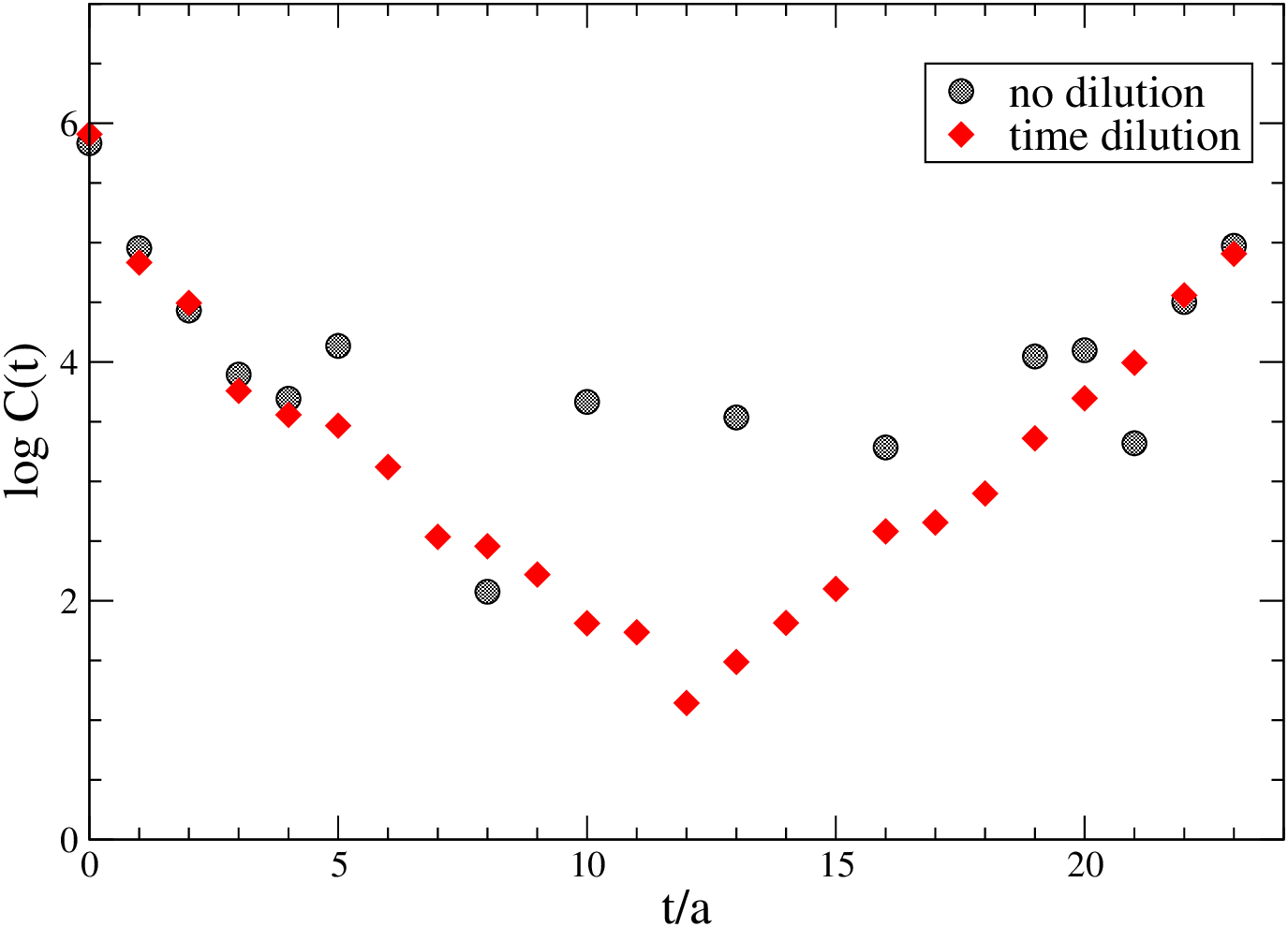}
\caption{The pseudoscalar propagator computed with and without time dilution.}
\label{fig:timedil}
\end{figure}

\begin{figure}
\includegraphics*[width=\colw]{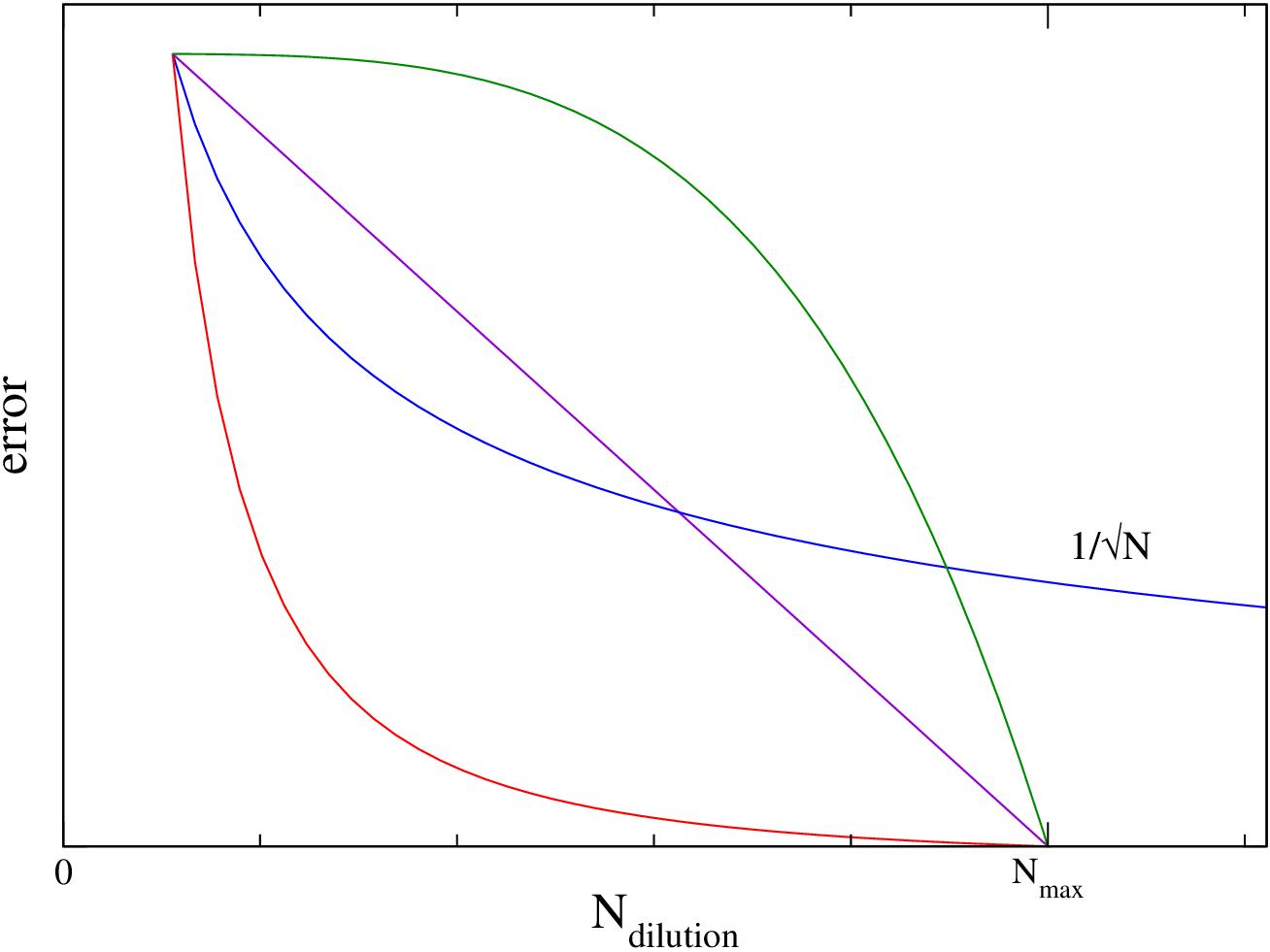}
\caption{A cartoon of possible deviations of the stochastic estimates
  of the exact solution (at $N_{dil}=N_{\text{max}}$) for different dilution
  paths.  Simply adding noise vectors will give a $1/\sqrt{N}$ behaviour. We have found that simple dilutions typically follow the behaviour exhibited by the bottom curve.}
\label{fig:cartoon}
\end{figure}

Each diluted source is inverted, yielding $N_d$ pairs of vectors,
$\{\psi^{(j)},\eta^{(j)}\}$, which then gives an
unbiased estimator of $M^{-1}$ with a {\em single} noise
source,
\begin{equation}
\sum_{i=0}^{N_d-1}\psi^{(i)}(\xv,t)\otimes\eta^{(i)}(\xv_0,t_0)^\dag \,.
\end{equation}
We show the effect of time dilution on a pseudoscalar propagator on a
$12^3\times24$ lattice in Fig.~\ref{fig:timedil}. The circles are
the average of 24 noise sources without any dilution and the diamonds
are from a single noise source which has been time-diluted.  This
particular scenario is analogous to the ``wall source on every
timeslice" method used by the authors of Ref.~\cite{Fukugita:1994ve}
to estimate the disconnected diagrams appearing in hadronic scattering
length calculations. Our method is, however, more general and can be
extended to the spin, colour and space components of the source
vector.\footnote{A dilution scheme that includes spin and colour, but
not time dilution, was previously used in Ref.~\cite{Wilcox:1999ab}.} The
``homeopathic'' limit of the dilution procedure, where we
have one noise vector for each time, space, colour and spin component, results
in the {\em exact} all-to-all propagator in a finite number of steps,
because of the property of Eq.~(\ref{eq:mod1}), see
Fig.~\ref{fig:cartoon}. This limit cannot be reached in practice on
realistic lattices, but the path of dilution may be optimised so that
the noise from the gauge fields dominate the errors in the hadronic
quantities of interest with only a small, manageable number of fermion
matrix inversions.

\subsection{Spectral Decomposition}
In the following we will make use of the hermitian Dirac operator
$Q\equiv \gamma_5 M$, where $M$ is the
usual Dirac operator. The quark propagator is $M^{-1}=Q^{-1}\gamma_5$, where $Q^{-1}$ is given
by a sum over all eigenmodes,
\begin{equation}\label{eq:spectral-exact}
Q^{-1}(y,x) =\sum_{i}^{N}\frac{1}{\lambda_i}v^{(i)}(y)\otimes
 v^{(i)}(x)^\dag \,.
\end{equation}
Here, $Qv^{(i)}=\lambda_iv^{(i)}$ and $N$ is the rank of the
matrix.  Computing all these eigenvectors is clearly not feasible;
however, theoretical arguments, supported by numerical evidence
\cite{Venkataraman:1997xi,Neff:2001zr,Duncan:2001ta}, suggest that much of the important 
infrared physics in hadronic interactions is encoded in the low-lying
eigenmodes.  Truncating the sum in Eq.~(\ref{eq:spectral-exact}) at some
finite $N\to\Nev$ should thus yield a good estimate of the quark
propagator for practical purposes.  Such a truncation does, however,
violate reflection positivity, making it mandatory to correct it.

\subsection{Hybrid Method}
Given the preceding discussion, a natural suggestion would be to try
to calculate as many
as possible of the low modes exactly and correct the truncation
with the noisy method. This gives rise to two concerns: firstly, the
correction should leave the exactly solved low-lying modes intact; and
secondly, it should not introduce large uncertanties in the
process. We propose that the stochastic method with noise dilution is
a natural way to accommodate both of those concerns.

First, we note that the exact $\Nev$ low eigenmodes obtained
separately naturally divide $Q$ into two subspaces, $Q=Q_0+Q_1$,
defined by
\begin{align}
Q_0&=\sum_{i=1}^{\Nev}\lambda_iv^{(i)}\otimes v^{(i)\dagger} \,,\qquad
Q_1=\sum_{j=\Nev+1}^{N}\lambda_jv^{(j)}\otimes v^{(j)\dagger} \,.
\end{align}
Similarly, the quark propagator is broken up into two pieces,
%
$Q^{-1}=\bQ_0+\bQ_1$,
where $\bQ_0$ is the truncated version of
Eq.~(\ref{eq:spectral-exact}) and $\bQ_1 = Q^{-1}\P_1$, 
where $\P_1$ is the projection operator
\begin{equation}
\P_1={\bf 1}-\P_0={\bf 1}-\sum_{j=1}^{\Nev}v^{(j)}\otimes v^{(j)\dagger}\,.
\end{equation}
We correct the truncation and estimate $\bQ_1$ using the stochastic method,
$\bQ_1=\bbra\psi_{[r]}\!\otimes\!\eta_{[r]}^\dagger\kket$
with $N_r$ noise vectors, $\{\eta_{[1]},\cdots,\eta_{[N_r]}\}$. The
solutions are given by
\begin{equation}\label{eq:remain}
\psi_{[r]}=\bQ_1\eta_{[r]}=Q^{-1}\left(\P_1\eta_{[r]}\right)\,.
\end{equation}
We now apply the idea of dilution to the
stochastic estimation of $\bQ_1$. Each random noise vector,
$\eta_{[r]}$, that is generated will be diluted and orthogonalised
(with respect to ${v^{(i)}}$) so that it can be used to obtain $\psi_{[r]}$. In
other words, we now have the following set of noise vectors:
\begin{equation}
\left\{\left(\eta_{[1]}^{(1)},\cdots,\eta_{[N_r]}^{(1)}\right),\cdots,\left(\eta_{[1]}^{(N_d)},\cdots,\eta_{[N_r]}^{(N_d)}\right)\right\}\notag
\end{equation}
where the upper indices denote the dilution and the lower indices
label the different noise samples. Note that the noise vectors are
mutually orthogonal due to the dilution before an average
over different random vectors are taken, i.e.,
\begin{equation}
\eta_{[r]}^{(i)}(\xv,t)\otimes\eta_{[s]}^{(j)\dagger}(\vec{y},t^\prime)=0
\quad\text{for all} \,i\neq j\,.
\end{equation}
This results in smaller variance than the standard method which mixes
noise, as Eq.~(\ref{eq:mixing}) shows.

\begin{figure}
\vspace{-4.5mm}
\includegraphics*[width=\colw]{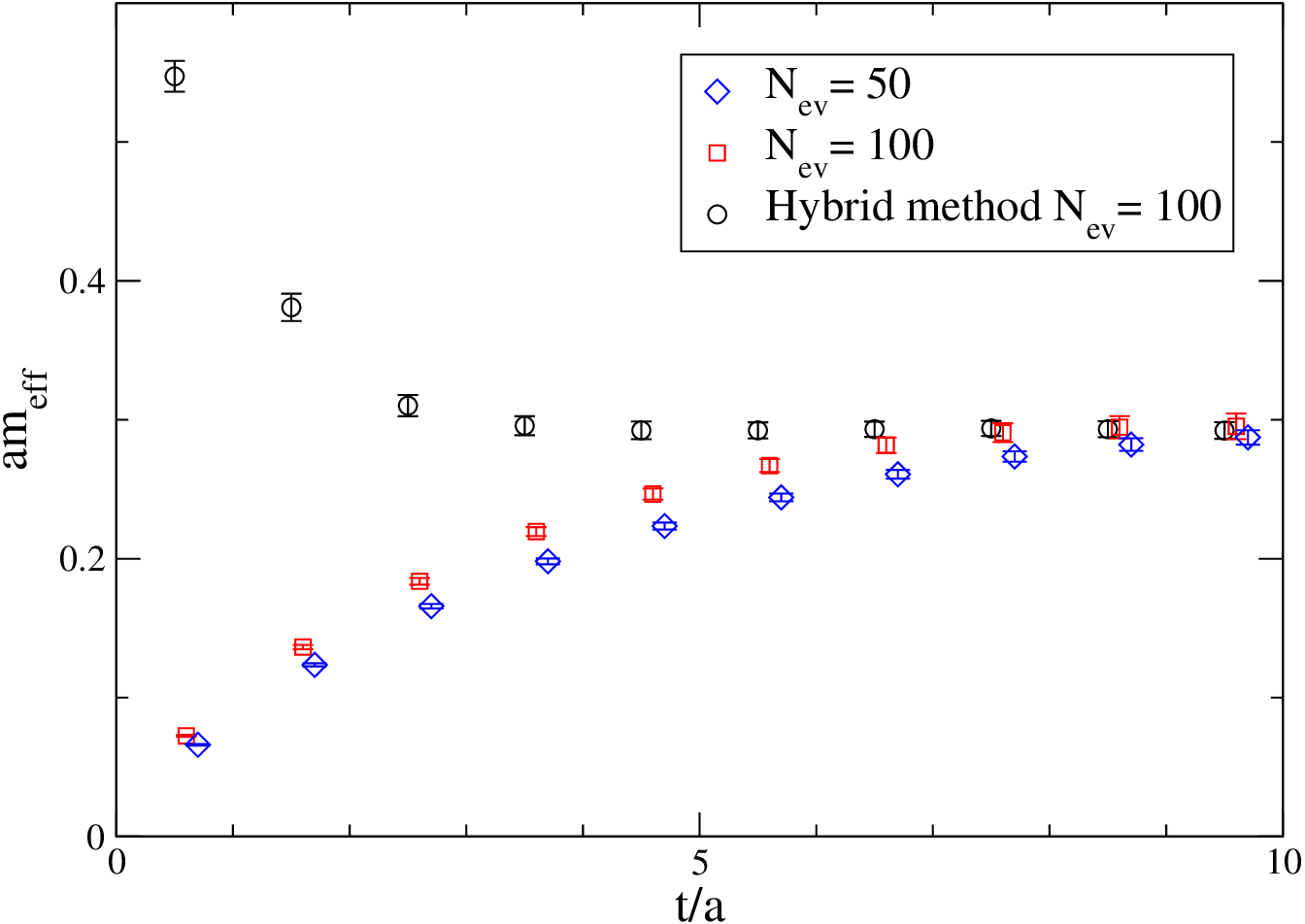}
\caption{The pion effective mass from 50, 100 eigenvectors and from the hybrid method with 100 and a time-diluted noise vector.}
\label{fig:hybrid-pi}
\end{figure}

There is a natural way to combine the two methods to estimate the
all-to-all quark propagator. The similarity in the structure of
Eq.~(\ref{eq:spectral-exact}) and Eq.~(\ref{eq:noisy}) suggests that one
construct the following ``hybrid list'' for the source and solution
vectors:
\begin{align}
w^{(i)}&=
\biggl\{\frac{v^{(1)}}{\lambda_1},\cdots,\frac{v^{(\Nev)}}{\lambda_{\Nev}},
\P_1\eta^{(1)},\cdots,\P_1\eta^{(N_d)}\biggr\}\label{eq:hybrid-w}\\
u^{(i)}&=
\biggl\{v^{(1)},\cdots,v^{(\Nev)},\psi^{(1)},\cdots,\psi^{(N_d)}\biggr\}
\label{eq:hybrid-u}
\end{align}
where the indices run over $N_{HL}=\Nev+N_d$ elements.
The unbiased, variance reduced estimate of the
all-to-all quark propagator (for a single random noise vector) is then given by
\begin{equation}
\sum_{i=1}^{N_{HL}}u^{(i)}(\xv,x_4)\otimes
 w^{(i)}(\vec{y},y_4)^\dagger\gamma_5\,.
\end{equation}
Note that defining $w^{(i)}$ without the projector $\P_1$ on the
$\eta$-vectors in Eq.~(\ref{eq:hybrid-w}) also gives an unbiased
estimator, which may however have a different variance.
Using the pion as an example, we can demonstrate how positivity is
recovered from the truncated propagator. In Fig.~\ref{fig:hybrid-pi}, we
show the effective mass from the truncated propagator and from the
hybrid method with a time, spin, colour and space (even-odd) diluted
noise vector. The truncated propagator, yielding an effective mass that approaches the
asymptotic value from below, is corrected by the addition of the
diluted noisy propagator.

%% file: implement.tex

Besides expanding the range of applications accessible in lattice QCD,
all-to-all propagators make the construction of hadronic interpolating
operators considerably simpler. With the (local) meson creation
operator
\begin{equation}
\Op_\Gamma(\xv,t) = \psibar(\xv,t)\Gamma\psi(\xv,t)\,,
\end{equation}
a meson propagator is constructed using traditional, point-to-all
propagators as follows,
\begin{equation}
C_\Gamma(\xv,t;\vec{0},0)=\Tr M^{-1}(\xv,t;\vec{0},0)\Gamma\gamma_5
 \;M^{-1}(\xv,t;\vec{0},0)^\dagger\gamma_5\bar{\Gamma}\,.
\end{equation}
This construction, where the local operators are traded for non-local
objects constructed from the quark propagators, may work well as long
as we restrict ourselves to operators localised at a single lattice
site.  However, an extended source operator involves
connecting quark propagators from sites in the vicinity of the source
site.  This requires at least an additional inversion
and entangles the matrix inversion and operator construction stages:
to a certain extent, the matrix inversion requires knowledge of the
operators to be used.

All-to-all
propagators eliminate this complication as both source and sink
operators are constructed purely from local vectors,
\begin{equation}
\Op_\Gamma^{(i,j)}(\xv,t)=w^{(i)}_{[1]}(\xv,t)^\dag\gamma_5\Gamma u^{(j)}_{[2]}(\xv,t)\,.
\end{equation}
The extra factor of $\gamma_5$ comes from the use of the
hermitian Dirac matrix, $\gamma_5M$.  Two pairs of vectors,
$(w_{[1]},u_{[1]})$ and $(w_{[2]},u_{[2]})$, are needed for the two
independent noisy estimators of the 
two propagators.  $\Gamma$ can now denote an
extended operator acting on the vector $u$.
For complicated operators such as those used to project out
hybrid or exotic states, this is a much needed simplification. For
example, an interpolating operator for the exotic hybrid $1^{-+}$ can
be constructed from combinations of gluonic paths projecting out the
relevant quantum numbers \cite{Lacock:1996vy}. One term in such a sum
may be
\begin{equation}
w^{[i]}(\xv)^\dag U_z(\xv)U_y(\xv+\hat{e}_z)U^\dag_z(\xv+\hat{e}_y)u^{[j]}(\xv+\hat{e}_y)\,,
\end{equation}
where the common $t$-index is suppressed.
A standard P-wave state may for example be constructed using
\begin{equation}
\Op^{(i,j)P}(\xv,t)=w^{(i)}(\xv,t)^\dag\gamma_5(D_ku^{(j)})(\xv,t)\,,
\end{equation}
where 
\begin{equation}
D_ku^{(j)}(\xv,t)=U_k(\xv,t)u^{(j)}(\xv+\hat{e}_k,t)-u^{(j)}(\xv,t)\,.
\end{equation}

Using all-to-all propagators, correlation functions are also
constructed in an intuitive manner. 
Hadronic correlation functions are obtained from
interpolating operators sitting at different time slices, e.g.
\begin{align}
 C^{AB}(\delta t) &= \sum_t\sum_{i,j}^{N_{HL}}
 {\Op}_{[1,2]}^{(i,j)A}(t)
 {\Op}_{[2,1]}^{(j,i)B}(t+\delta t)\,,\label{eq:isovec-prop}\\
\intertext{where}
 {\Op}_{[1,2]}^{(i,j)A}(t) 
 &= \sum_{\xv}w^{(i)}_{[2]}(\xv,t)^\dag\gamma_5
 \Gamma^A u^{(j)}_{[1]}(\xv,t)
 \equiv \bra w^{(i)}_{[2]}(t),\gamma_5\Gamma^A u^{(j)}_{[1]}(t)\ket
\label{eq:corr_op_field}\\
\intertext{and}
 {\Op}_{[2,1]}^{(j,i)B}(t+\delta t)
  &= \bra w^{(j)}_{[1]}(t+\delta t),\gamma_5
 \Gamma^B u^{(i)}_{[2]}(t+\delta t)\ket\,,
\end{align}
for isovector two-point correlators. The correlator is constructed as
a product of these meson operators on different time slices.  The
operators $\Gamma^{A,B}$ may now also include phases for momentum
projections.  As a consequence, once the hybrid list vectors have been
constructed, correlation functions for any operator can be computed
without any additional inversions.

\subsection{Noise recycling}

We can take advantage of
having saved different random source/solution samples by reusing them
in the contraction. In other words, one can generate $N_R$ samples of
noise vectors, $\eta_{[r]}$, save the corresponding solutions,
$\psi_{[r]}$ to disk and perform the contraction,
\begin{equation}
C^{AB}(t,t_0)=\sum_{r<s}\bra w_{[r]}(t),\Gamma^Au_{[s]}(t)\ket
 \bra w_{[s]}(t_0),\Gamma^Bu_{[r]}(t_0)\ket \,,
\end{equation}
yielding $\sim\!N_R^2$ samples of the correlation
function. The errors correspondingly decrease faster than the naive
$1/\sqrt{N_R}$, although the measurements are somewhat correlated. We
have seen in our preliminary tests that the error reduction is
comparable to some dilution choices. It is clear that if one can
afford to save the noise and solution vectors onto disk, then this is a
straightforward method of variance reduction for mass-degenerate mesons.

\subsection{Improving Performance}

When constructing one of these hadron operators, such as Eq.~(\ref{eq:corr_op_field}), on a particular time slice $t$ and for a particular operator $A$, the hybrid technique can be manipulated for computational efficency.

Firstly, compute (and store)
\begin{equation}
z^{(j)}_{[1]}(\xv,t) = \gamma_5\Gamma^A u^{(j)}_{[1]}(\xv,t),\, \forall j \label{eq:req_op}.
\end{equation}
What now remains for the calculation of the time-slice elements of (Eq.~\ref{eq:corr_op_field}) is $N_{HL}\times N_{HL}$ dot-products of
spinors on a time slice, since
\begin{equation}
\Op_{[1,2]}^{(i,j)A}(t)
 = \sum_{\xv}w^{(i)}_{[2]}(\xv,t)^\dag z^{(j)}_{[1]}(\xv,t) 
\equiv \bra w^{(i)}_{[2]}(t), z^{(j)}_{[1]}(t)\ket \,.
\end{equation}
We note that time dilution is taken as a minimum requirement in the case of meson two-point functions 
and that the $w^{(i)}$ hybrid list is naturally divided between our two vector spaces, since
\begin{equation}
w^{(i)}=
\biggl\{\frac{v^{(1)}}{\lambda_1},\cdots,\frac{v^{(\Nev)}}{\lambda_{\Nev}},\P_1\eta^{(1)},\cdots,\P_1\eta^{(N_d)}\biggr\}\,.
\end{equation}
If we evaluate (and store)
\begin{equation}\label{eq:store}
\bra w^{(i)}_{[2]}(t), z^{(j)}_{[1]}(t)\ket
=\frac{1}{\lambda_i}\bra v^{(i)}_{[2]}(t), z^{(j)}_{[1]}(t)\ket \quad ,\forall i\le \Nev, \forall j\,,
\end{equation}
the elements that remain to be computed are
\begin{equation}\label{eq:operator}
 \bra\eta^{(i)}_{[2]}(t), ({\bf 1}-\P_0) z^{(j)}_{[1]}(t)\ket
=\bra\eta^{(i)}_{[2]}(t), z^{(j)}_{[1]}(t)\ket -
\sum_{k=1}^{\Nev}\alpha^{(ik)}_{[22]}
 \bra v^{(k)}_{[2]}, z^{(j)}_{[1]}(t)\ket \,,
\end{equation}
where
\begin{equation}
\alpha^{(ik)}_{[22]} = \bra\eta^{(i)}_{[2]}(t), v^{(k)}_{[2]}\ket
 = \bra\eta^{(i)}_{[2]}, v^{(k)}_{[2]}\ket \,.
\end{equation}
The first term only needs to be computed if $\eta^{(i)}_{[2]}(t)$ has
support on the relevant time slice. The second term is simply a
weighted sum over terms from the eigenvector space of $w^{(i)}$ that
have already been calculated. The weights are given by {\em global} dot
products of quark vectors which can be calculated externally to any
particular time slice or operator: Since each $\eta$ has support on
only a single time slice, the global dot product is equal to the dot
product of $v$ with the time-slice restricted $\eta(t)$.

This method also reduces the amount of storage required for the noise
quarkfields by a factor of $N_t$ since only the supported time slice
needs to be stored on disk, all others being identically zero.
This argument is easily extended to include the conjugate operator 
and all required momenta.

Once the general routine to construct a operator field
$\Op^{(i,j)A}_{[k,l]}(t)$, as in Eq.~(\ref{eq:corr_op_field}), and the
summation over the hybrid list indices required in
Eq.~(\ref{eq:isovec-prop}) are in place, the method becomes a black box
to the end user. From Eq.~(\ref{eq:req_op}), we see that this user
need only supply the subroutines to create the required hadron
operators from the quark, antiquark and gluon fields on a time slice
for each of the required operator fields. These routines would only
contain the operations to be performed on the quark field, such as
$D_j\psi$, and have no reference to the hybrid lists.

For isoscalar mesons, the disconnected part of the propagator must be
included, yielding the following contraction,
\begin{multline}
C^{(I=0)}_\Gamma(t,t_0)
=\bra w^{(i)}_{[2]}(t),\gamma_5\Gamma u^{(j)}_{[1]}(t)\ket
\bra w_{[1]}^{(j)}(t_0),\gamma_5\Gamma^\dag u_{[2]}^{(i)}(t_0)\ket\\
-\bra w_{[1]}^{(j)}(t),\gamma_5\Gamma u_{[1]}^{(j)}(t)\ket
\bra w_{[2]}^{(i)}(t_0),\gamma_5\Gamma^\dag u_{[2]}^{(i)}(t_0)\ket\,.
\end{multline}

\subsection{Recipe to compute two-point functions}

\begin{enumerate}
\renewcommand{\labelenumi}{\theenumi.}

\item Calculate $\Nev$ eigenvectors, $v_{j}$, of
$\gamma_{5}M$ with eigenvalues $\lambda_{j}$.

\item Calculate solution vectors, $\psi^{(i)}$, of
\begin{equation}
\gamma_{5}M\psi^{(i)} = \P_1\eta^{(i)}, \, \forall i
\end{equation}
where $\eta^{(i)}$ are diluted noise vectors, satisfying
Eqs~(\ref{eq:mixing}), (\ref{eq:mod1}) and (\ref{eq:timedilute}) and 
\begin{equation}
\P_1={\bf 1}-\P_0={\bf 1}-\sum_{j=1}^{\Nev}v^{(j)}\otimes v^{(j)\dagger}\,.
\end{equation}
At least two independent
noise vectors need to be generated for the two quark propagators required
in two-point functions.

\item With the hybrid lists defined as in Eqs~(\ref{eq:hybrid-w}) and
(\ref{eq:hybrid-u}), construct the (time-slice) vectors
$z^{(j)}_{[r]}(t)$,
\begin{equation}
z^{(j)}_{[r]}(t) = \gamma_5\Gamma^A u^{(j)}_{[r]}(t), \forall j.
\end{equation}
where the operation $\Gamma^A u^{(j)}_{[r]}(t)$ is defined by the end
user, and can also represent a conjugate operation.

\item Construct the operator field $\Op_{[1,2]}^{(i,j)A}(t)$
\begin{equation}
\Op_{[1,2]}^{(i,j)A}(t) = \bra w^{(i)}_{[2]}(t), z^{(j)}_{[1]}(t)\ket \,,
\end{equation}
on a time slice $t$ and store it to disk. Repeat for all time slices and for both the source and sink operators.

\item Contract these operator fields to obtain the two-point
correlator $C^{AB}$,
\begin{equation}
C^{AB}(\delta t)  =  \sum_t\sum_{i,j}^{N_{HL}}
 \Op_{[1,2]}^{(i,j)A}(t) \Op_{[2,1]}^{(j,i)B}(t+\delta t) \,.
\end{equation}

\item If increased accuracy is required, calculate $\Nev^{\hphantom{ev}\prime} - \Nev$ additional eigenvectors of
$\gamma_{5}M$. The solution vectors
$\psi^{(i)}$ can be projected into the reduced orthogonal eigenvector space in
$\gamma_{5}M$ by the addition of a correction term,
\begin{equation}
\psi^{(i)} \rightarrow \psi^{(i)} - \sum_{j=\Nev}^{\Nev^{\hphantom{ev}\prime}}
\frac{1}{\lambda_{j}} \bra v_{j},\eta^{(i)}\ket v_{j}
\end{equation}
One can also increase the dilution level, from $N_d$ to
$N_{d}^{\prime}$ diluted noise vectors, with only $N_{d}^{\prime} -
N_d$ extra quark inversions. For example, if
%
$\gamma_{5}M\psi = \P_1\eta $
%
with $\eta = \eta^{(1)} + \eta^{(2)}$, 
we need only calculate the solution to 
\begin{equation}
\gamma_{5}M\psi^{(1)} = \P_1\eta^{(1)} 
\end{equation}
since $\eta^{(2)} = \eta - \eta^{(1)}$
and $\psi^{(2)} = \psi - \psi^{(1)}$.

\end{enumerate}

%% file: results.tex

In this exploratory study, we use a data set consisting of 75
quenched configurations at $\beta=5.7$ on a $12^3\times24$ lattice.
We have used Wilson fermions with hopping parameter $\kappa=0.1675$,
corresponding to $m_\pi/m_\rho=0.50$ \cite{Butler:1994em}.
The 100 lowest eigenvectors have
been computed for each configuration, and noise vectors have been
diluted in time, spin, colour and in space on an even/odd basis. 

\subsection{Isovector mesons}

We have computed correlators for the pion, rho, the $1^{+-}, 0^{++}, 1^{++}$
and $2^{++}$ P-wave states, using spatially extended operators for the
P-waves \cite{Lacock:1996vy}. The quark
fields were smeared with 10 levels of Jacobi smearing iterations. In cases
where we use multiple operators for a single state, we have found some
differences in the overlap depending on the spatial size of the
operators where typically larger operators were favoured. 

\begin{figure}
\includegraphics*[width=\colw]{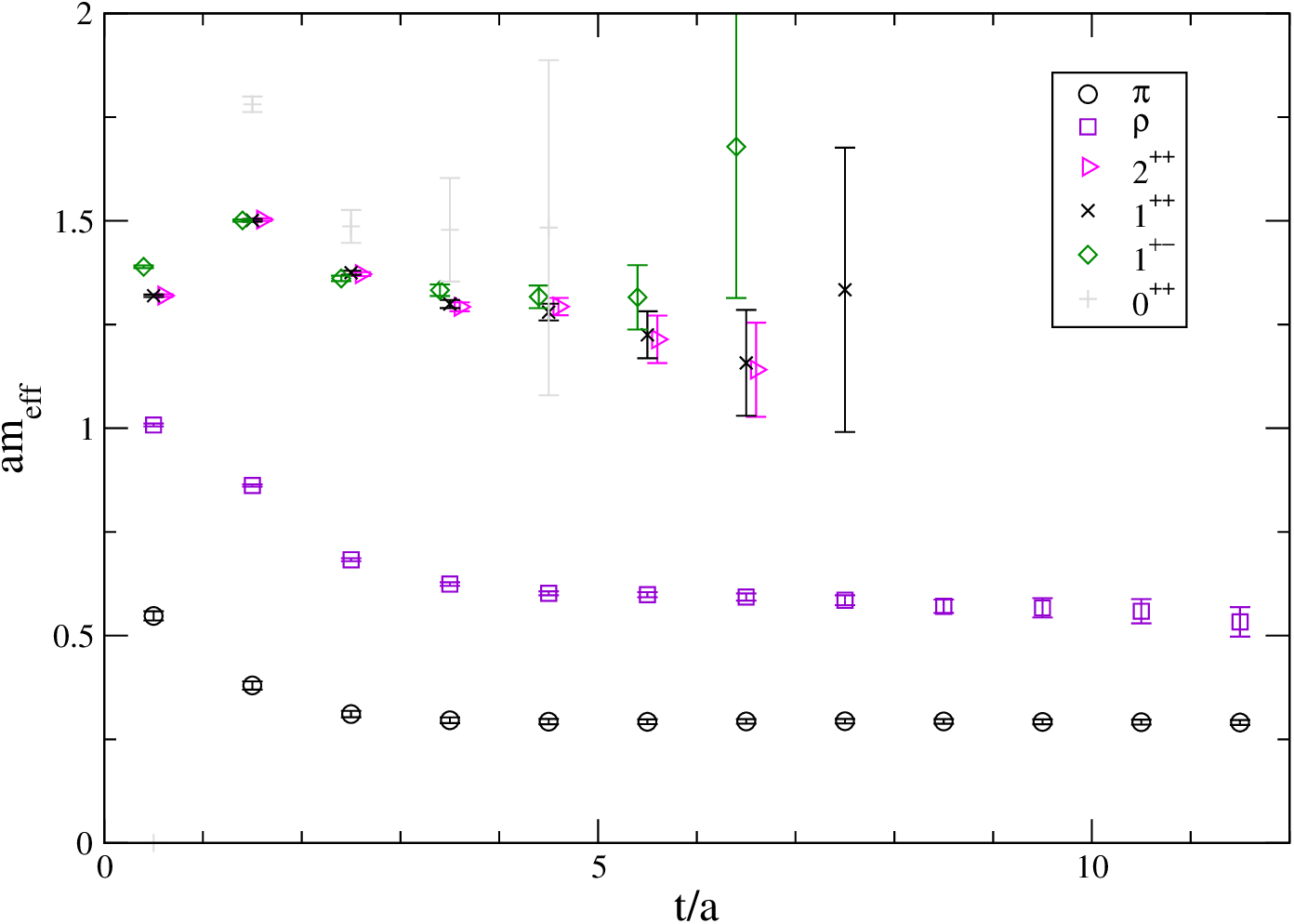}
\caption{Effective masses for isovector mesons with 100 eigenvectors,
  time, colour, spin and space-even-odd dilution and 75 configurations.}
\label{fig:spectrum}
\end{figure}

Figure \ref{fig:spectrum} shows effective masses for the pion, rho and
P-wave states, using 100 eigenvectors, time, colour, spin and
space-even-odd dilution and 75 configurations. A lower statistics
result for the hybrid $1^{-+}$ was reported in Ref.~\cite{O'Cais:2004ww}.

\begin{figure}
\includegraphics*[width=\colw]{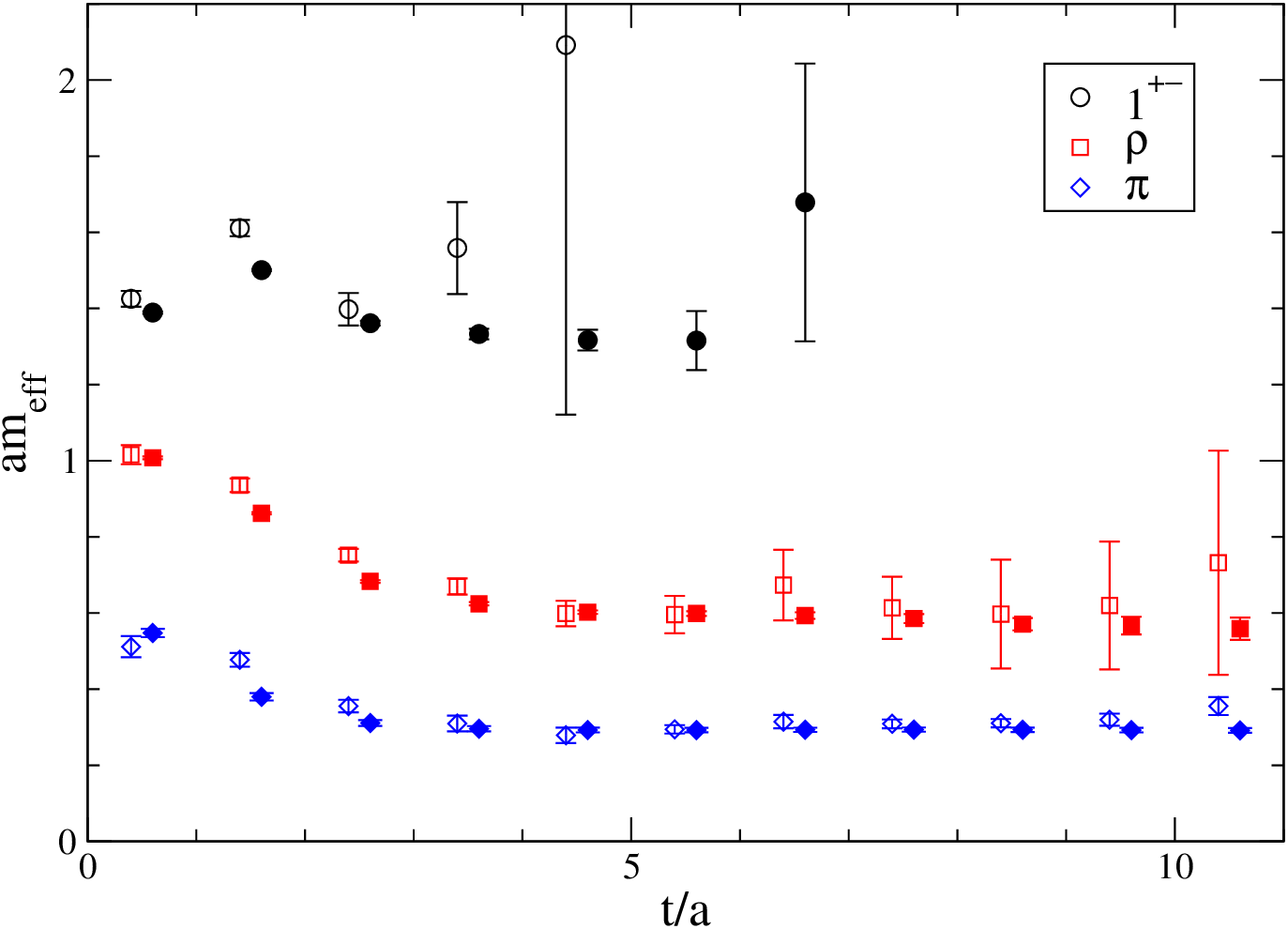}
\caption{Comparison of spectrum from all-to-all and point-to-all
  propagators.  The open symbols are point-to-all, the closed symbols
  all-to-all with 100 eigenvectors and time, colour, spin and 
	space-even-odd dilution.}
\label{fig:pointcompare}
\end{figure}

We can also compare the results using our method with results using
traditional point-to-all propagators on the same number of
configurations.  These are shown in Fig.~\ref{fig:pointcompare} for
the pion, rho and $1^{+-}$ P-wave. The P-wave was computed using point
sources at 3 different spatial points in order to construct the
derivative operator. While the improved signal for the
pion may not justify the additional computational expense, it is clear
that for the rho meson and in particular the P-wave the reduction in
noise is very significant with all-to-all propagators.  This is
particularly important if only a limited number of configurations is
available, which will be the case for light, dynamical full QCD configurations.  
In the case of P-waves (as well as hybrids, and of course
isoscalar mesons) all-to-all propagators may be the only way of
obtaining any acceptable signal at all.

\subsubsection{Effect of dilution}

\begin{figure}
\includegraphics*[width=\colw]{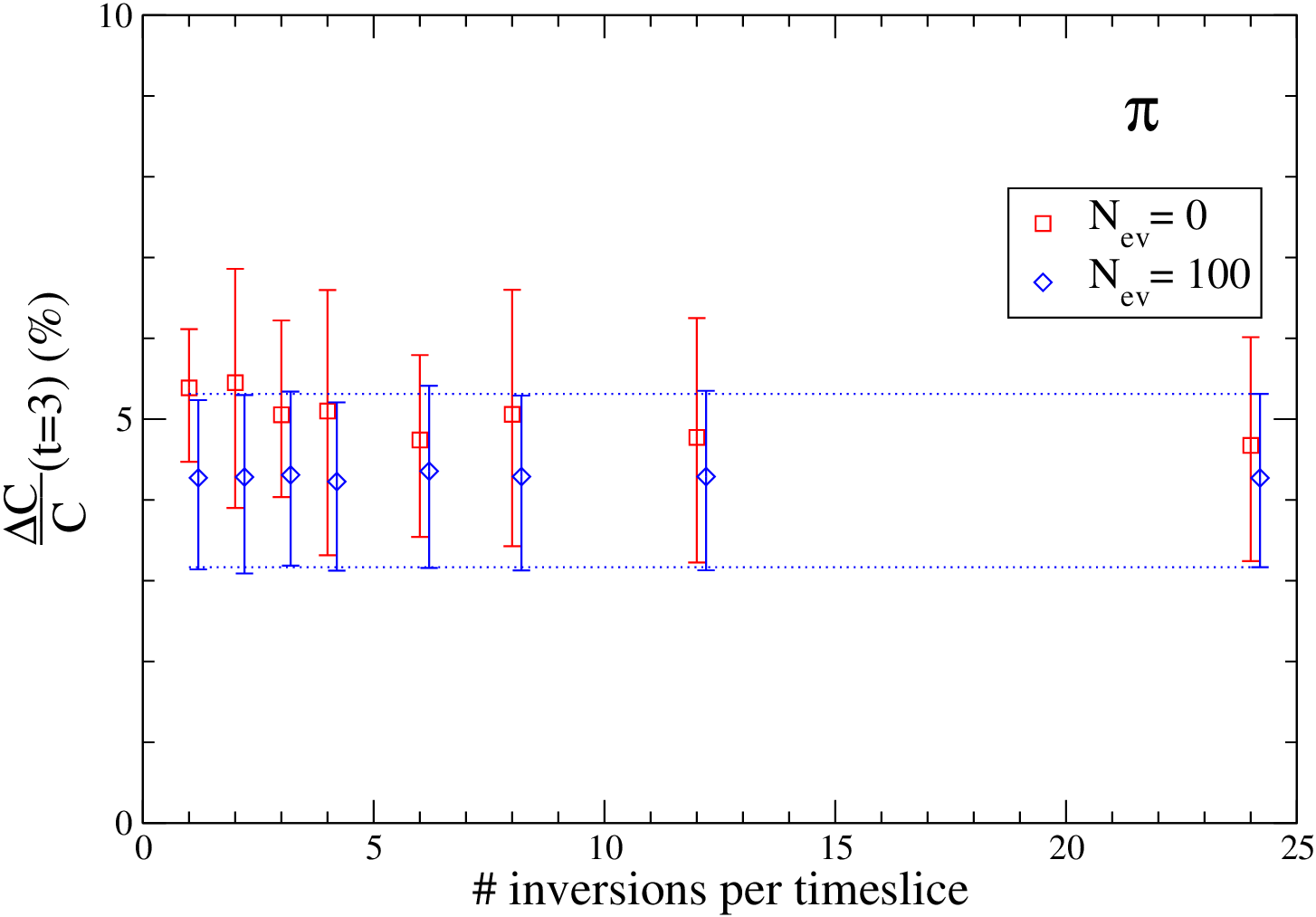}
\caption{Demonstration of the effect of dilution on the statistical errors as
  a function of dilution, for the pion.  The squares denote the errors
  obtained using only noise vectors, while the diamonds are errors
  obtained using the hybrid method.}
\label{fig:error-pion}
\end{figure}

\begin{figure}
\includegraphics*[width=\colw]{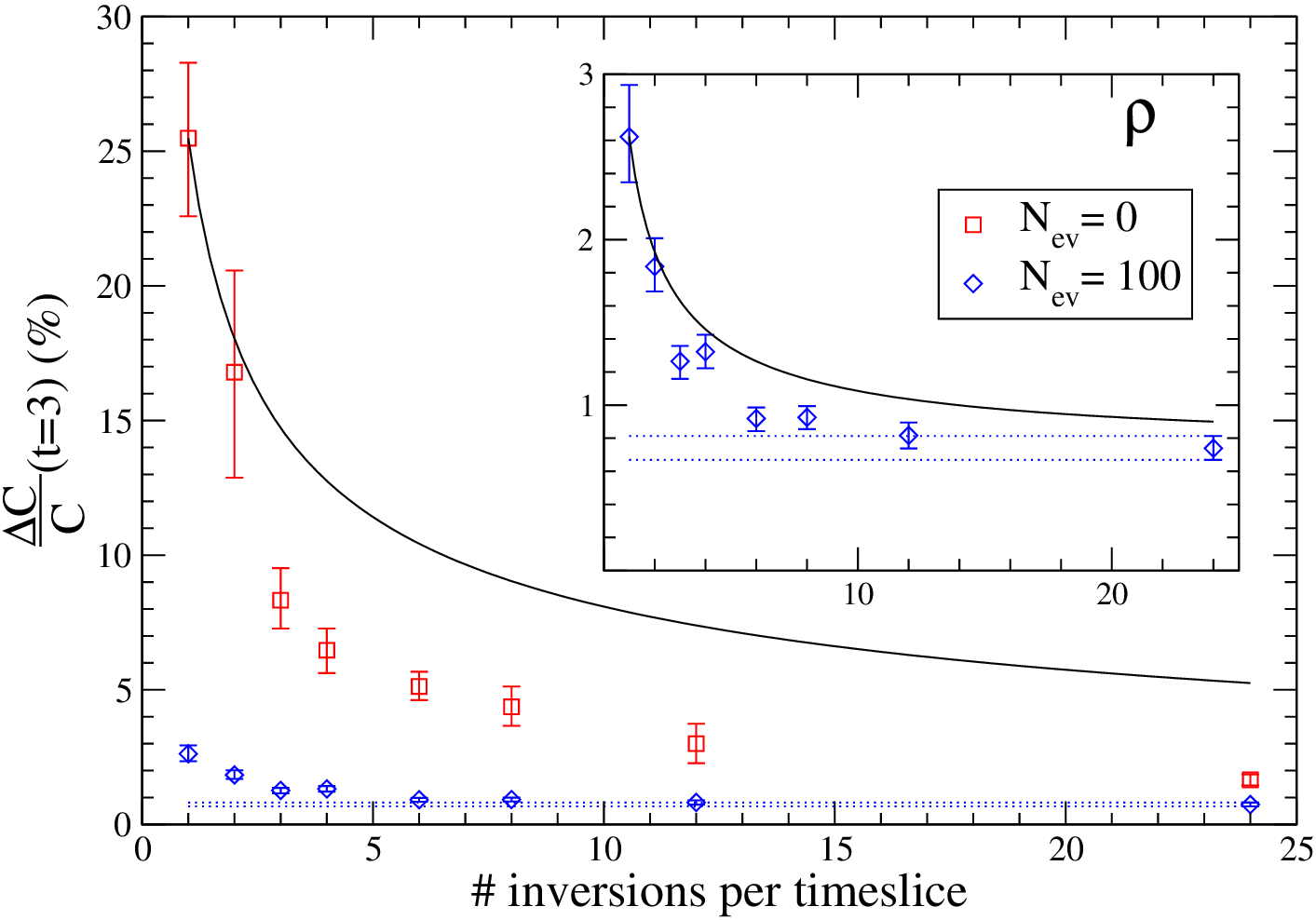}
\caption{Demonstration of the effect of dilution on the statistical errors as
  a function of dilution, for the $\rho$.  The squares denote the
  relative errors on the correlator on time slice 3 obtained using
  only noise vectors, while the
  diamonds are errors obtained using the hybrid method.  The solid
  line is the expected behaviour of the errors from accumulating an
  equivalent number of independent (time-diluted) noise samples.  The 
  inset shows the relative errors from the hybrid method in more detail.}
\label{fig:error-rho}
\end{figure}

\begin{figure}
\includegraphics*[width=\colw]{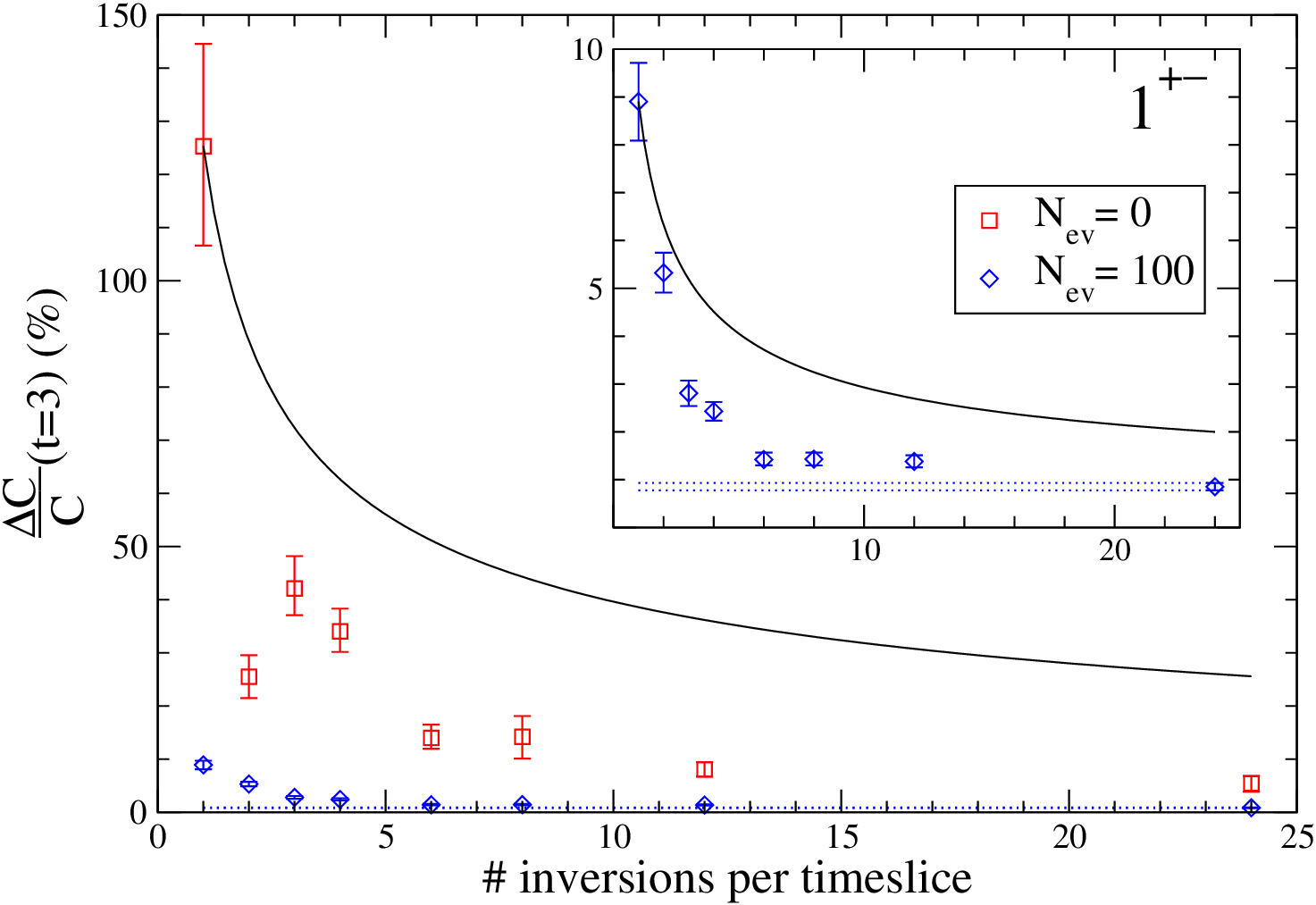}
\caption{As Fig.~\ref{fig:error-rho}, for the $1^{+-}$ P-wave.}
\label{fig:error-1pm}
\end{figure}

\begin{figure}
\includegraphics*[width=\colw]{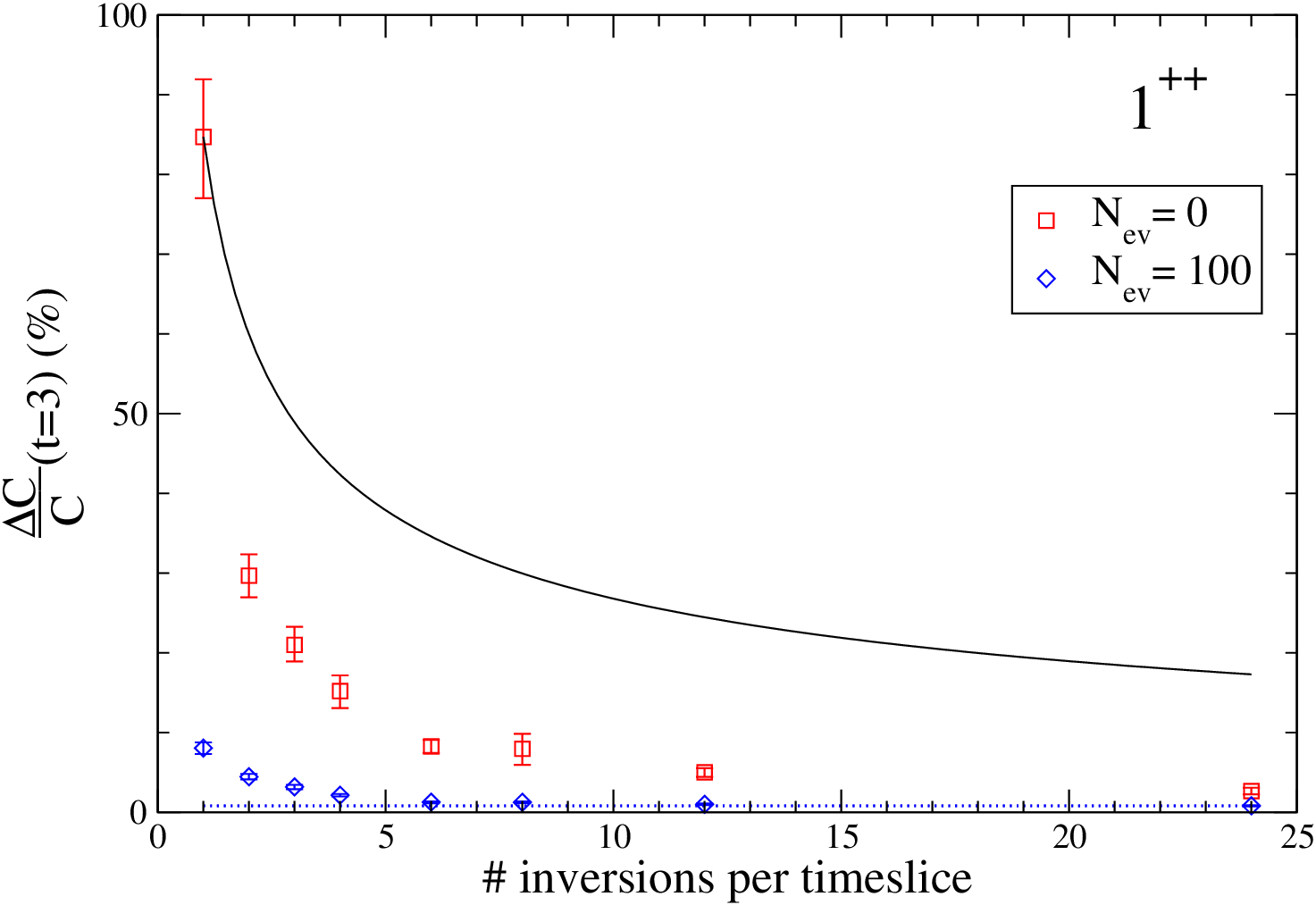}
\caption{As Fig.~\ref{fig:error-rho}, for the $1^{++}$ P-wave.}
\label{fig:error-1pp}
\end{figure}

\begin{figure}
\includegraphics*[width=\colw]{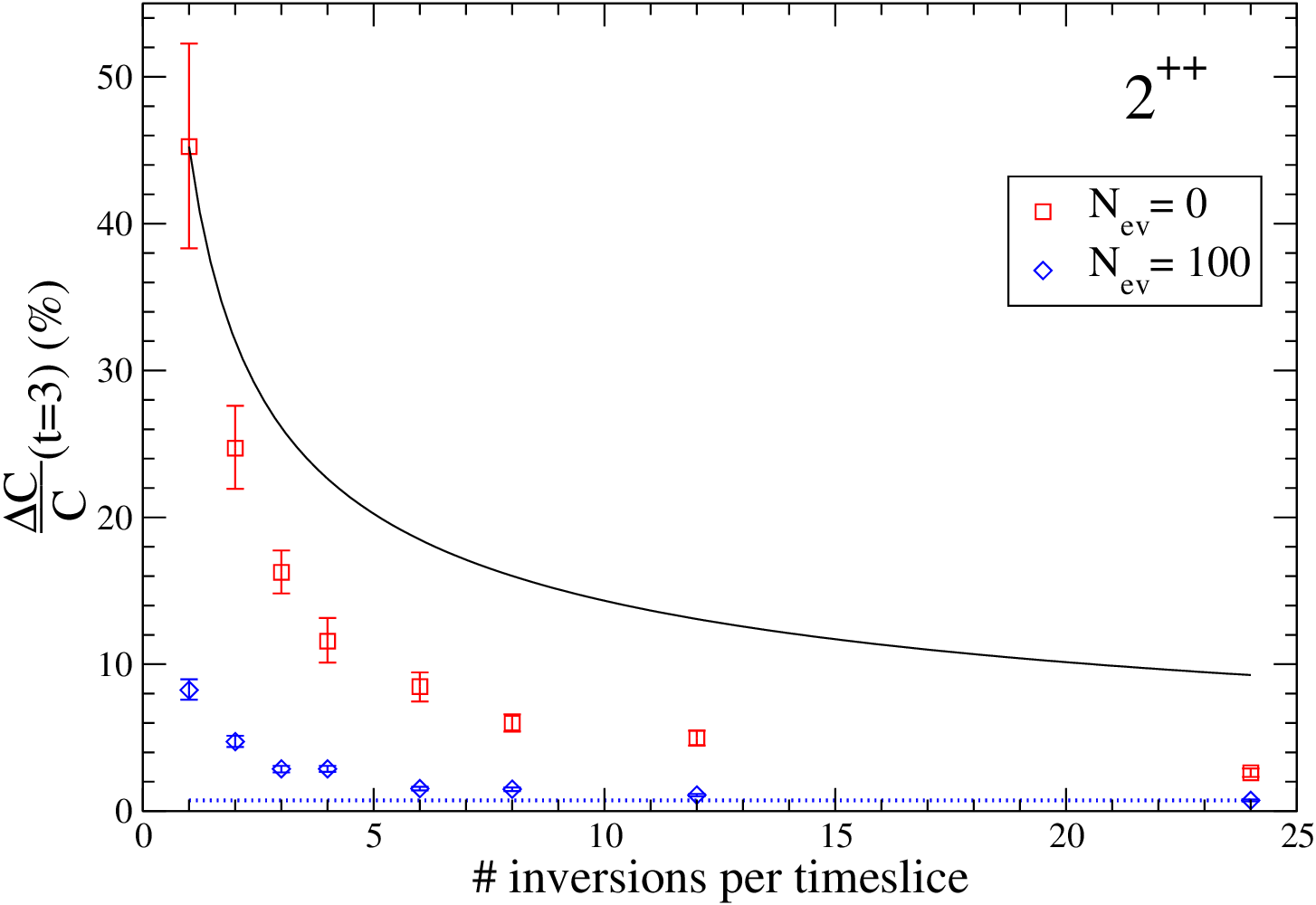}
\caption{As Fig.~\ref{fig:error-rho}, for the $2^{++}$ P-wave.}
\label{fig:error-2pp}
\end{figure}

First we study the effect of dilution using only noise vectors and no
eigenvectors. Results for the pion, rho, $1^{+-}$, $1^{++}$ and
$2^{++}$ are shown in Figures~\ref{fig:error-pion}-\ref{fig:error-2pp}
with square symbols. The error bars on the errors were obtained by
bootstrap resampling.  Time-dilution is sufficient to reach the level
where the error on the correlator is saturated by the gauge fluctuations for the
pion. This is not the case for the vector as the error decreases with
increasing number of dilutions.  We also show the curve which would be
expected to result from using $N$ independent noise samples per
time slice.  This curve is given by
\begin{equation}
\sigma_N^2 = \frac{\sigma_1^2-\sigma_g^2}{N} + \sigma_g^2\,,
\end{equation}
where $\sigma_g$ is the gauge noise, i.e. the statistical uncertainty
resulting from having a finite number of gauge configurations.  We
have assumed that the
gauge noise is given by the errors from the hybrid method with our
highest dilution level, $\sigma_g=\sigma_{24}^{\text{hyb}}$.  We
see in all cases studied here that dilution always works better than
accumulating an equivalent number of independent noise samples.  A
closer inspection reveals that the errors fall off like $1/N$ rather
than the naive $1/\sqrt{N}$. This is in addition to the exponential
gain due to time dilution.

\subsubsection{Effect of eigenvectors}

\begin{figure}[t]
\includegraphics*[width=\colw]{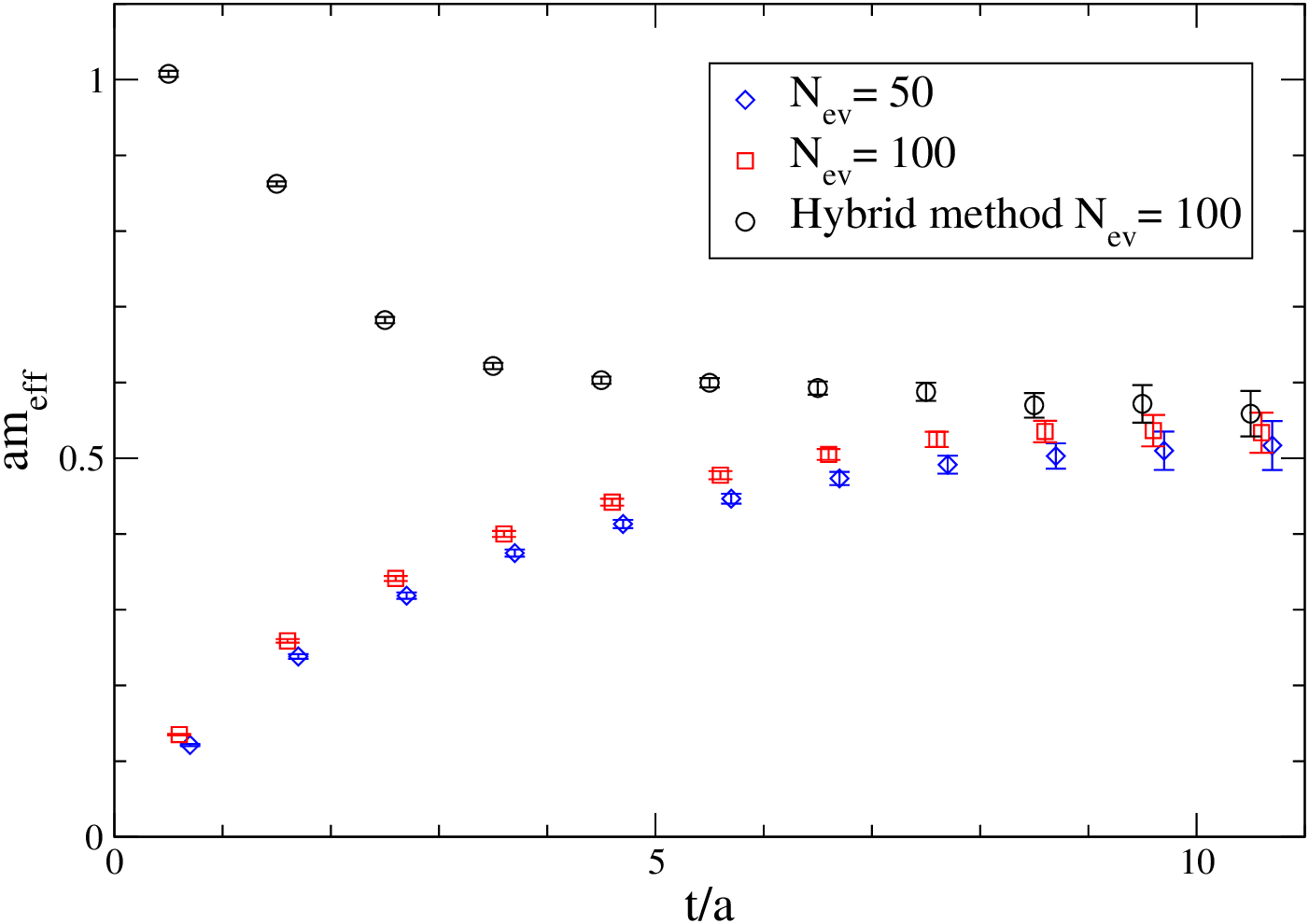}
\caption{The $\rho$ effective masses from the 50 and
  100 lowest eigenvectors alone, and from the hybrid method with 100 
	eigenvectors. The corresponding figure for the pion is shown in 
	Fig.~\ref{fig:hybrid-pi}.}
\label{fig:hybrid-rho}
\end{figure}

Figures~\ref{fig:hybrid-pi} and ~\ref{fig:hybrid-rho} shows how the 
low-lying eigenvectors saturate the signal for the pion and to a lesser 
extent the rho meson.  For early times the effective mass from the 
eigenvectors alone clearly exhibits violation of positivity, but from $t=7$ 
onwards for the pion and $t=9$ for the rho, the full signal for the effective 
mass is captured by the eigenvectors.  We also note the clear difference 
between the data from 50 and 100 eigenvectors. We have not performed any 
systematic study of the effect of varying the number of eigenvectors in 
the hybrid method, since this is expected to be highly dependent on the 
action and lattice spacing used, and the non-chiral action and coarse 
lattices used in this exploratory study would be unlikely to yield valuable 
information for more realistic simulations.

The diamonds (lower points) in Figs
\ref{fig:error-pion}--\ref{fig:error-2pp} show the
behaviour of the errors for the pion, rho, $1^{+-}$, $1^{++}$ and
$2^{++}$ as a function of the dilution level for the hybrid method
with 100 eigenvectors.  We observe a dramatic decrease in the relative
error of the estimated correlator when the lowest 100 eigenvectors are used.
The exception is the pion, where the errors remain constant at about
5\% (from our 75 configurations) regardless of dilution and
eigenvectors.  We interpret this as an effect of the pathologies of
Wilson fermions on coarse lattices with light quarks, where the large
fluctuations of the lowest eigenmode affect the pion in particular.

The insets of Figs \ref{fig:error-rho} and \ref{fig:error-1pm} show the errors
for the hybrid method in more detail as a function of dilution, for
the $\rho$ meson and $1^{+-}$ P-wave.  The
gauge noise level appears to be reached somewhere between dilution
levels 6 (colour and space even/odd) and 12 (colour and spin), although a
slight downward trend may still be observed at $N_{\text{dil}}=24$ for
the P-waves. It is interesting to note that fractional error reached for the
vector meson is much smaller than the pseudoscalar (for this rough
action).

\subsection{Isoscalar mesons}

\begin{figure}
\includegraphics*[width=\colw]{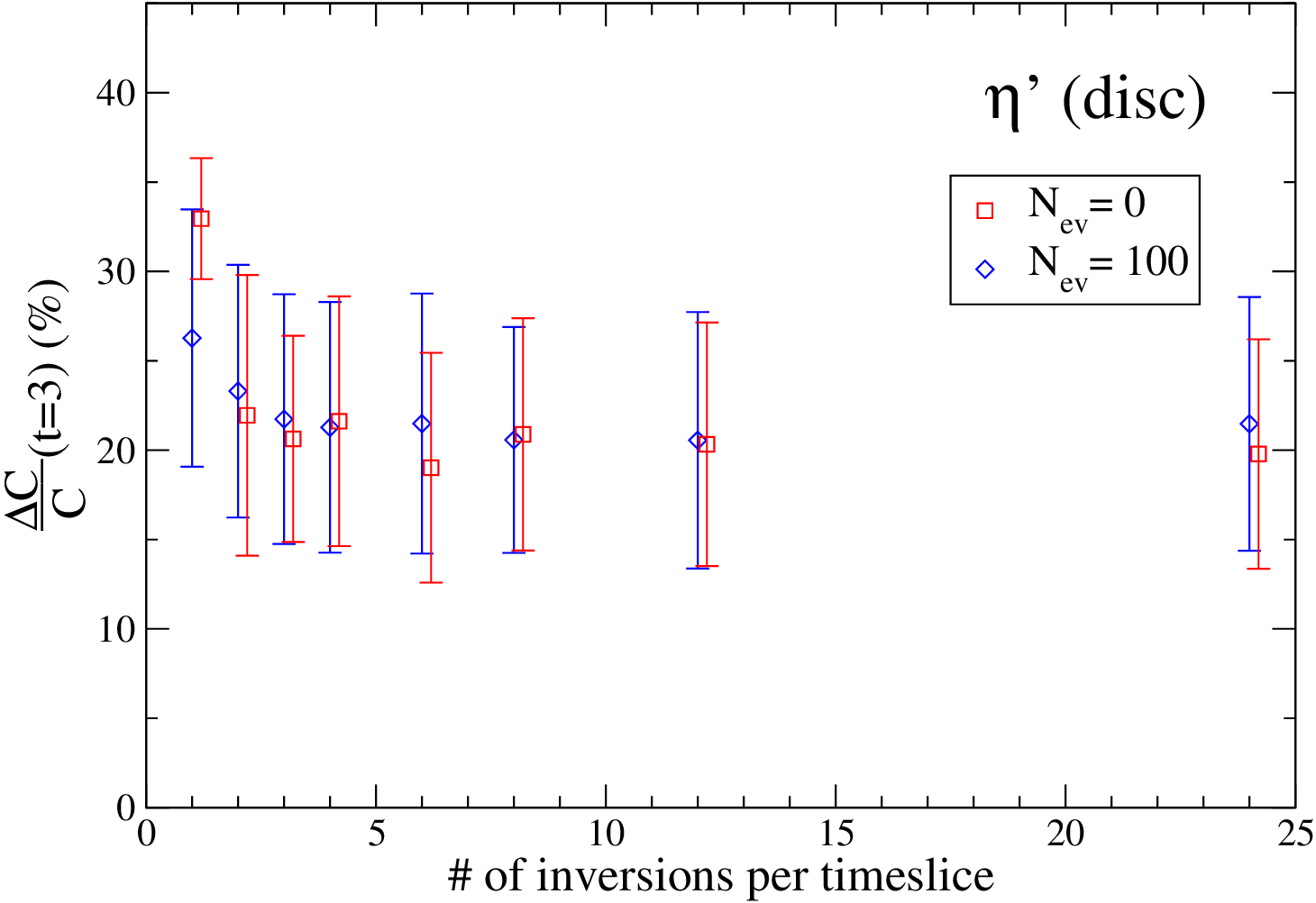}
\caption{As fig.~\ref{fig:error-rho}, for the disconnected piece
of the isoscalar correlation function.}
\label{fig:error-eta-disc}
\end{figure}

We have computed the disconnected part of the pseudoscalar isoscalar
($\eta'$) meson.  Figure~\ref{fig:error-eta-disc} shows the errors as
a function of dilution level for the hybrid method ($\Nev=100$) and
using only a noisy estimator.  We see that the errors from the gauge
noise are large, of the order of 20\%, and consequently the total
uncertainty is saturated by the gauge noise almost immediately.  This
is not surprising on this coarse lattice with a non-chiral action, but
it gives confidence that a good signal can be obtained from a proper
action.

\subsection{Static--light mesons}
The benefit of using all-to-all propagators is particularly 
evident in simulations of heavy-light mesons in the static limit. 
Using a single point propagator for the light quark means that 
the source and sink of the static-light correlator are restricted 
to a single spatial lattice site. All-to-all propagators 
allow us to place source and sink operators at each spatial site 
on the lattice, yielding a dramatic increase in statistics. The 
application of all-to-all propagators to the static-light 
spectrum has previously been examined in Refs.~\cite{Michael:1998sg,Green:2003zz}. 

The mesons considered in the previous sections are classified 
according to their $J^{P C}$ quantum numbers. Here, however,
the mesons contain non-degenerate quarks and charge conjugation is 
no longer a symmetry of the meson. Also, static-light mesons which 
differ only in the spin of the static quark are degenerate. 
Therefore, in the static limit, we find a single S-wave channel and 
two distinct P-wave channels. We label these $J_{\ell}^{P}$ where 
$J_{\ell}$ is the total angular momentum of the light degrees of freedom. 
The S-wave is then $\frac{1} {2}^{-}$ and the P-wave channels are $\frac{1} {2}^{+}$ 
and $\frac{3} {2}^{+}$. 

\begin{figure}
\includegraphics*[width=\colw]{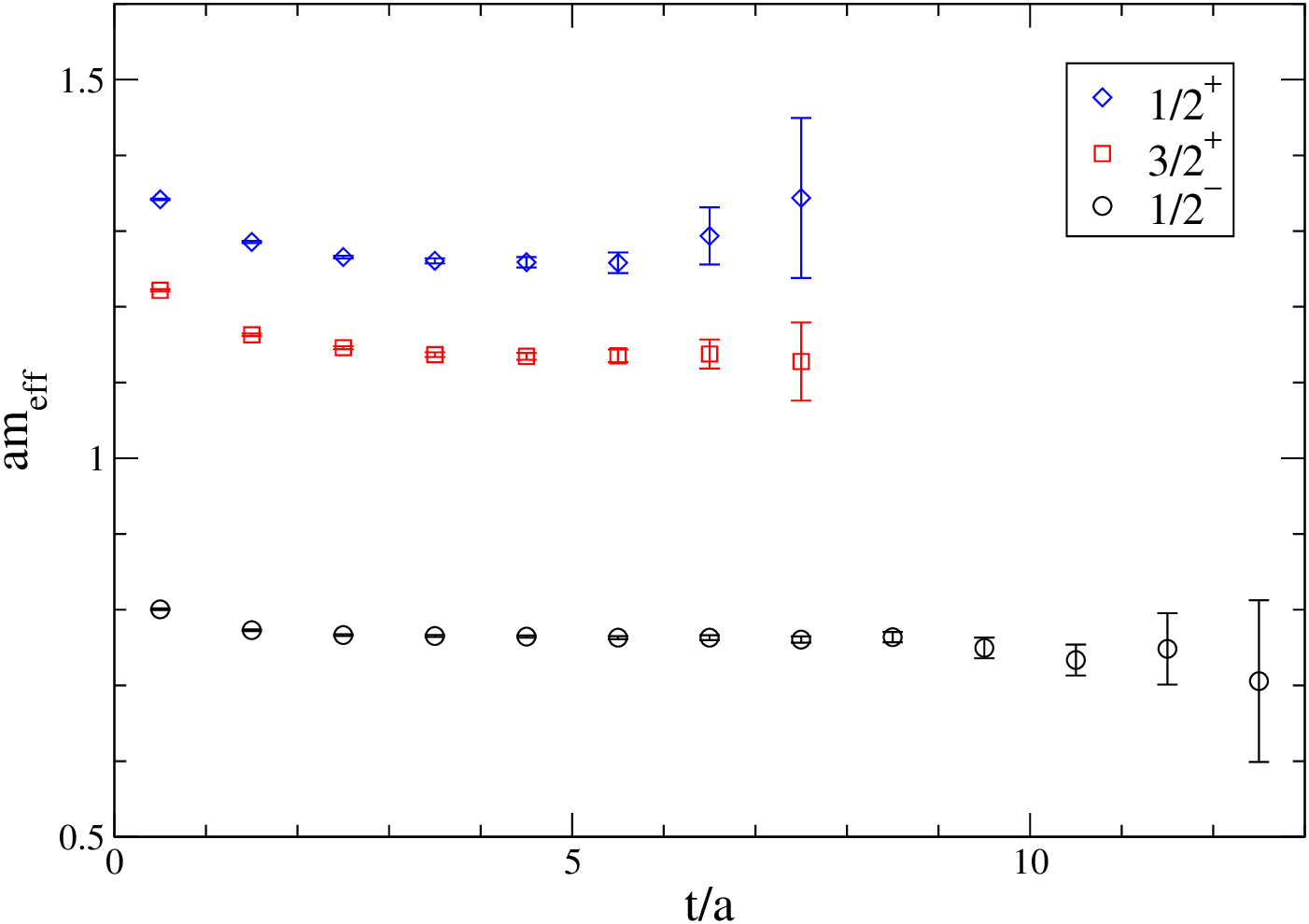}
\caption{Effective masses for the static--light S-wave and P-waves,
calculated using 100 eigenvectors and time, colour, spin and space-even-odd 
dilution.}
\label{fig::sl-full}
\end{figure}

\begin{figure}
\includegraphics*[width=\colw]{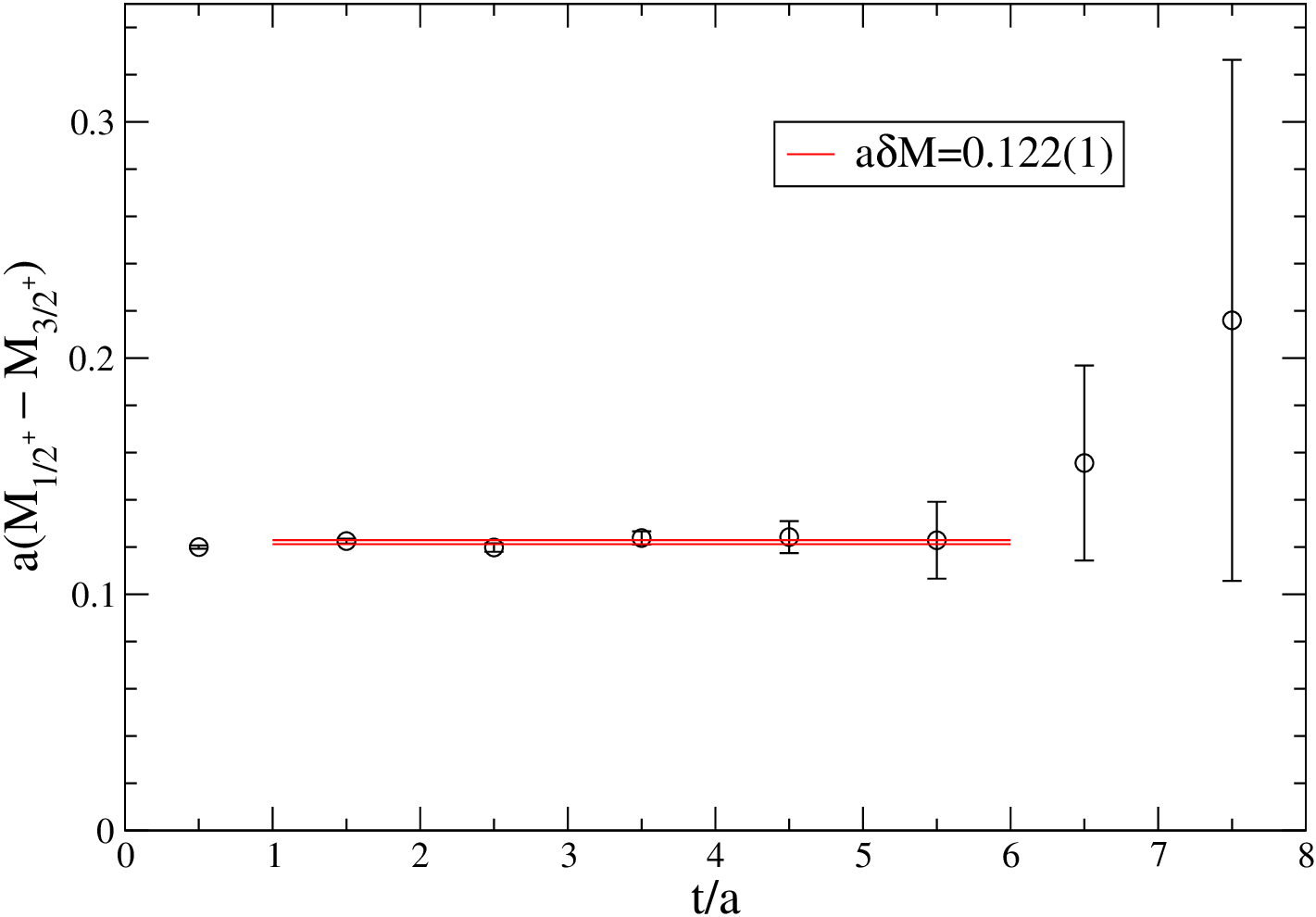}
\caption{The energy-difference between the static--light P-wave 
ground states, with the line denoting the best fit.}
\label{fig:sl-ediff}
\end{figure}

Figure~\ref{fig::sl-full} shows effective masses for the static-light 
mesons. The run parameters are the same as for the isovector mesons 
shown in Fig.~\ref{fig:spectrum} and, once again, we have used extended 
operators for the P-waves. We obtain excellent signals in each channel and, 
as Fig.~\ref{fig:sl-ediff} shows, we are able to determine the P-wave 
ground-state energies to within $1\%$ accuracy. Physically, the 
$\frac{3}  {2}^{+}$ is expected to be heavier than the $\frac{1} {2}^{+}$~\cite{Green:2003zz} but, given the coarseness 
of the lattice and the actions used, we do not expect our results to be 
physically meaningful.  

\begin{figure}
\includegraphics*[width=\colw]{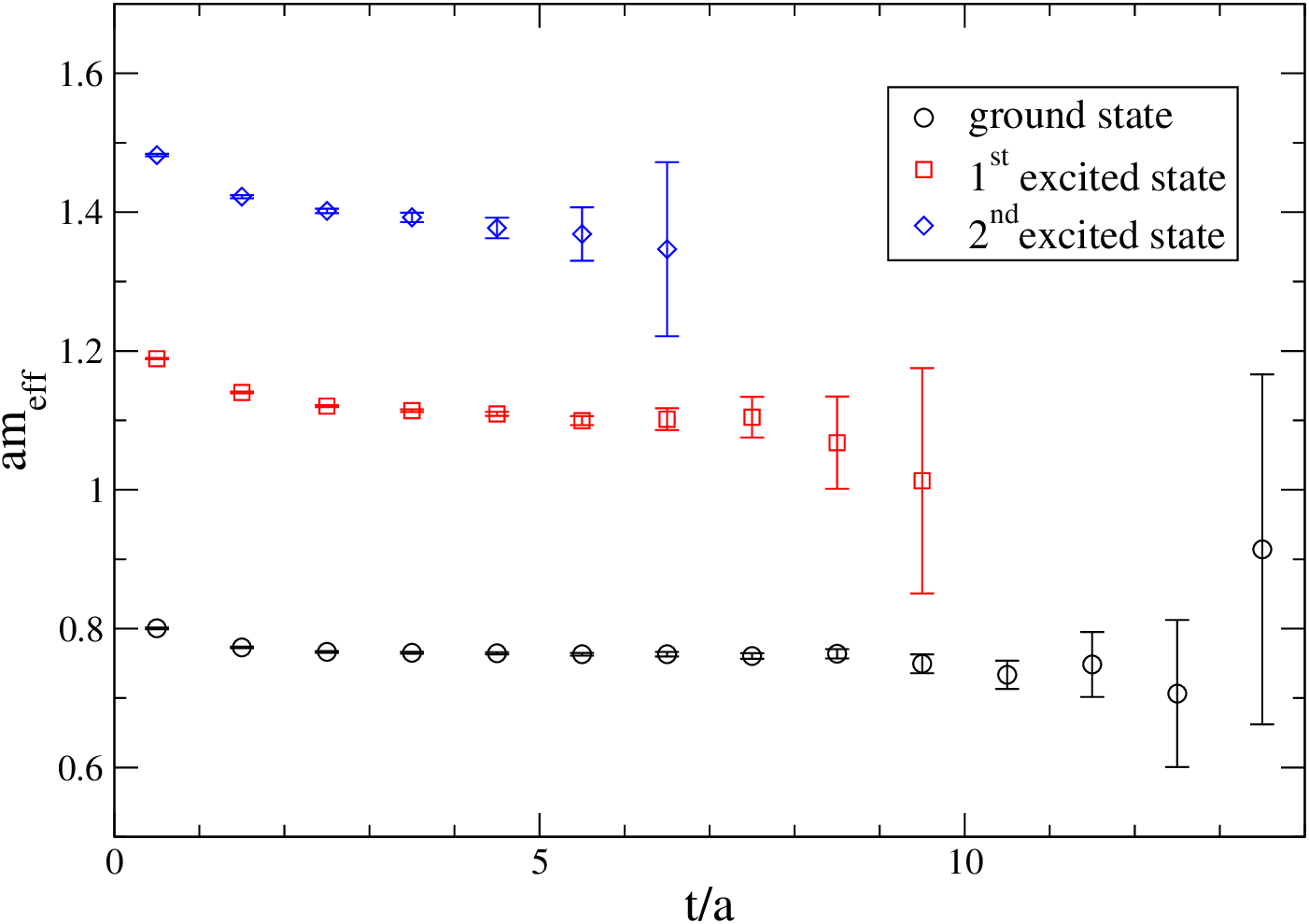}
\caption{Effective masses for the three lowest-lying states of the 
static--light S-wave.}
\label{fig:sl-excited}
\end{figure}

The results were obtained using the variational method with five operators 
differing by the Jacobi smearing parameters of the light quark. This is a 
very simple way of extending the basis operators to get a better overlap 
with the ground state and for identifying excited states. As 
Fig.~\ref{fig:sl-excited} shows, we find clear signals for the first and 
second excited states of the S-wave. We can also resolve signals for the 
first excited states of both P-waves.

\begin{figure}
\includegraphics*[width=\colw]{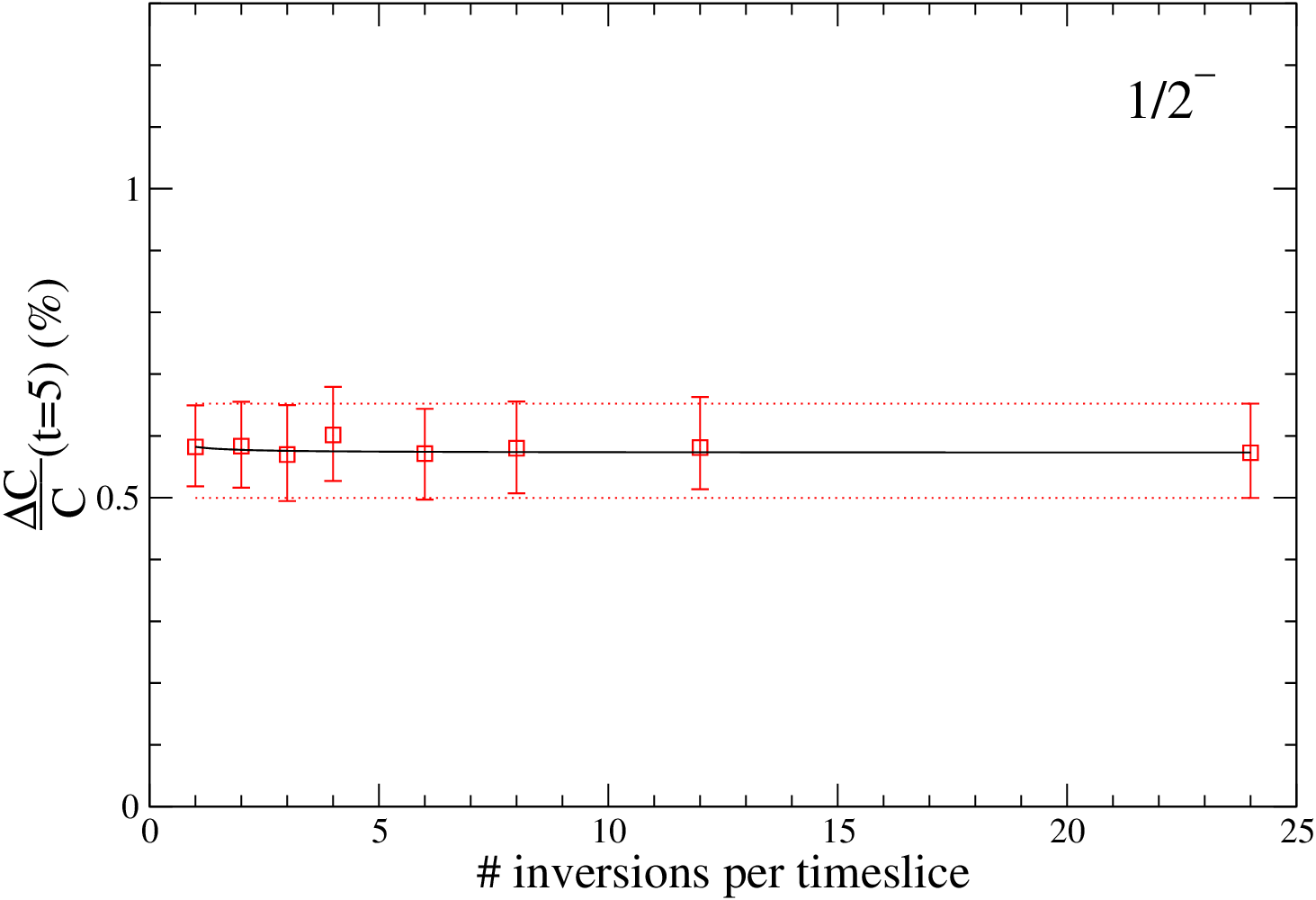}
\caption{Demonstration of the effect of dilution on the error-bars as
  a function of dilution, for the static--light S-wave, with 100 
  eigenvectors.  The solid line is the expected behaviour of the
  errors from accumulating an equivalent number of independent,
	time-diluted noise samples.}
\label{fig:error-SLswave}
\end{figure}

\begin{figure}
\includegraphics*[width=\colw]{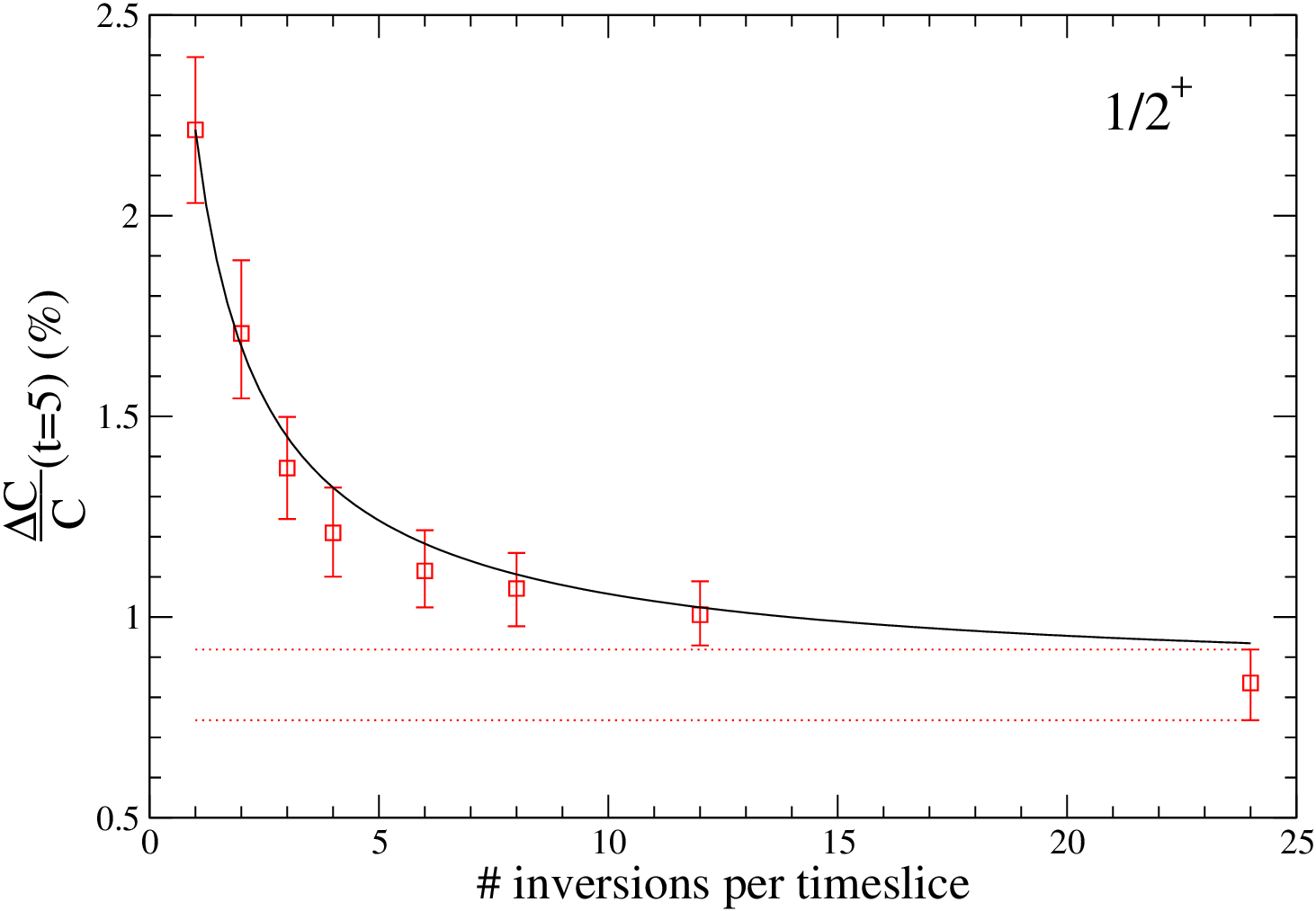}
\caption{As fig.~\ref{fig:error-pion}, for the static--light $\frac{1} {2}^{+}$ P-wave.}
\label{fig:error-SLG1}
\end{figure}

\begin{figure}
\includegraphics*[width=\colw]{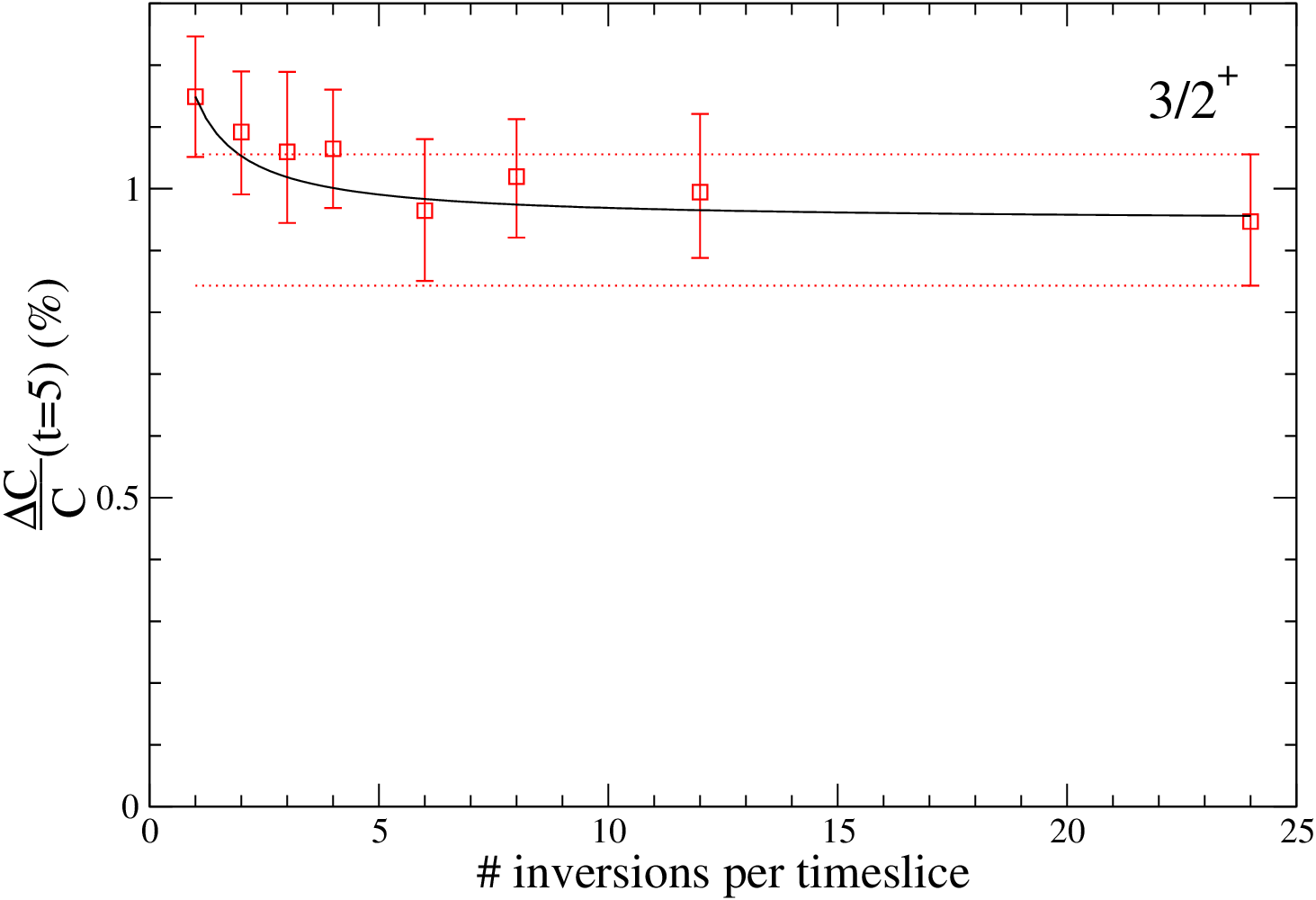}
\caption{As fig.~\ref{fig:error-pion}, for the static--light $\frac{3} {2}^{+}$ P-wave.}
\label{fig:error-SLH}
\end{figure}

Figures \ref{fig:error-SLswave}--\ref{fig:error-SLH} show the relative
errors for the static--light meson correlators as a function of
dilution level.  For the $\frac{1} {2}^-$ and $\frac{3} {2}^+$ the errors are already
saturated by the gauge noise with time dilution alone, while for the
$\frac{1} {2}^+$ there is still some gain to be had from additional dilution.  In
this case we find that the errors fall off approximately as
$1/\sqrt{N}$, indicating little if any sensitivity to the dilution
path for this variable: after time dilution, the purely statistical
gain is the dominant factor.

%% file: conclude.tex

We have presented a new algorithm to estimate the all-to-all
propagator.  All-to-all propagators make it possible to make use of
all the available information in a gauge configuration, which
considering the cost to generate full QCD configurations may be of
crucial importance. They are also a necessary ingredient in
flavour-singlet physics and in computing fermionic thermodynamic
quantities.

All-to-all propagators have a further advantage over point propagators
in that operator construction is considerably simplified: the
operators are constructed in a natural way from local fields, and
extended operators used in variational methods may be employed at no
additional cost.

We have presented evidence that diluting stochastic estimators in
time, colour, spin or other variables gives less variance than
traditional noisy estimators.  This is not unexpected, since dilution
will yield the exact all-to-all propagator in a finite number of
iterations.  More work is needed to determine the optimal dilution
path for different observables, although the simplest choices (colour,
space even/odd, and to some extent spin dilution) are often sufficient
to reach the gauge noise level for the ensemble sizes considered here.

The hybrid method allows one to extract the important physics from
low-lying eigenmodes and combine this with a noisy correction in a
natural way.  This may be implemented so that the user can
be blind to the details of the dilution and eigenvector list.

We have found that this method makes it possible to determine
traditionally ``noisy'' quantities such as masses of P-wave states,
hybrids and static--light mesons (including excited states) to a high
level of precision with limited statistics.

Although we have focused on (mainly isovector) mesons in this study,
the method is also straightforwardly applicable to baryons and to
thermodynamic quantities (condensates and susceptibilities).  Work on
flavour singlets and thermodynamic observables is currently in
progress.